\documentclass[12pt, a4paper]{article}

\providecommand{\keywords}[1]{\textbf{\textit{keywords: }} #1}
\DeclareUnicodeCharacter{2009}{\,} % Thinn space Unicode definition 
\newenvironment{acknowledgement}{%
  \renewcommand{\abstractname}{Acknowledgements}% Rename Abstract to Acknowledgements
  \begin{abstract}
}{%
  \end{abstract}
}

\usepackage{xr}
\makeatletter
\newcommand*{\addFileDependency}[1]{
  \typeout{(#1)}
  \@addtofilelist{#1}
  \IfFileExists{#1}{}{\typeout{No file #1.}}
}
\makeatother
\newcommand*{\myexternaldocument}[1]{
    \externaldocument{#1}
    \addFileDependency{#1.tex}
    \addFileDependency{#1.aux}
}
\myexternaldocument{supplement}
\usepackage{tikz}           % advanced plotter
\usepackage{adjustbox}      % Resizing plotter
\usepackage{bm}             % bold math, to use \boldsymbol
\usepackage{amsmath}        % Use \text in the equation
\usepackage{amsfonts}       % Use \mathbb in the equation

\usepackage{censor}            % Cencsoring
\StopCensoring                 % Stop

\title{Efficient discovery of multiple minimum action pathways using Gaussian process}

\author{\censor{JaeHwan Shim}$^1$\thanks{\texttt{schinavro@snu.ac.kr}}, 
        \and \censor{Juyong Lee}$^2$\thanks{\texttt{nicole23@snu.ac.kr}}, \and \censor{Jaejun Yu}$^1$\thanks{\texttt{jyu@snu.ac.kr}}}
\date{%
    $^1$\censor{\textit{Department of Physics and Astronomy, Seoul National University}}, \\
    \censor{\textit{1 Gwanak-ro, Gwanak-gu, Seoul 08826, Republic of Korea}} \\
    $^2$\censor{\textit{College of Pharmacy, Seoul National University}}, \\ 
    \censor{\textit{1 Gwanak-ro, Gwanak-gu, Seoul 08826, Republic of Korea}}\\[2ex]%
    \today
}

\usepackage[english]{babel}
\usepackage[square,numbers, sort&compress]{natbib}
\usepackage{hyperref}       % Hyperlink to packge
\hypersetup{
    colorlinks=true,    
    breaklinks=true,                
    urlcolor= black,                
    linkcolor= blue,                
    bookmarksopen=false,
    filecolor=black,
    citecolor=blue,
    linkbordercolor=blue,
    pdfborder={0 0 1}
}
\begin{document}

\maketitle

\begin{abstract}
We present a new efficient transition pathway search method based on the least action principle and the Gaussian process regression method.
Most pathway search methods developed so far rely on string representations, which approximate a transition pathway by a series of slowly varying system replicas. 
Such string methods are computationally expensive in general because they require many replicas to obtain smooth pathways.
Here, we present an approach employing the Gaussian process regression method, which infers the shape of a potential energy surface with a few observed data and Gaussian-shaped kernel functions.
We demonstrate a drastic elevation of computing efficiency of the method about five orders of magnitude than existing methods.
Further, to demonstrate its real-world capabilities, we apply our method to find multiple conformational transition pathways of alanine dipeptide using a quantum mechanical potential. %\textit{ab initio} calculations on that of 
Owing to the improved efficiency of our method, Gaussian process action optimization (GPAO), we obtain the multiple transition pathways of alanine dipeptide and calculate their transition probabilities successfully with \textit{ab initio} accuracy.
In addition, GPAO successfully finds the isomerization pathways of small molecules and the rearrangement of atoms on a metallic surface.
\end{abstract}

\keywords{Gaussian Process, Onsager Machlup}

% keywords can be removed

\section{Introduction}

Finding multiple transition pathways of phase transitions, chemical reactions, and conformational transitions remains challenging in physics and chemistry.
With conventional molecular dynamics (MD) or Monte-Carlo (MC) approaches, i.e., initial-value formulation, it takes a considerably long time to simulate an evolution over energy barriers starting from an initial state. 
%If energy barriers are high, it is not easy to observe such events even with state-of-the-art computers. 
To overcome such limitations, numerous single-end methods have been suggested~\cite{cerjan_finding_1981, broyden_quasi-newton_1967,doye1997surveying,munro_defect_1999,malek_dynamics_2000,ohno_scaled_2004, maeda_global_2005, laio_escaping_2002, barducci_metadynamics_2011, shang_stochastic_2013, prinz_markov_2011, broadbelt_computer_1994,broadbelt_computer_1994,matheu_mechanism_2003,gao_reaction_2016,zimmerman_automated_2013, maeda_exploring_2014}.
Methods include using Hessian to climb up the energy barrier~\cite{cerjan_finding_1981, broyden_quasi-newton_1967}, using the eigenvalues of a Hessian matrix~\cite{doye1997surveying, munro_defect_1999}, repeating the process of climbing up and relaxing to locate multiple local minima~\cite{malek_dynamics_2000}, and using the knowledge, which atoms are more likely to form or to break a bond, to guide the transition~\cite{broadbelt_computer_1994, matheu_mechanism_2003, gao_reaction_2016,maeda_exploring_2014}, were suggested.
The referred methods find the transition state efficiently.
However, these methods do not guarantee finding a transition pathway reaching a targeted product state. 
%However, it lacked on specifying a pathway will system have on transition, or estimate its properties.

Another class of approaches, biased sampling methods, were introduced to sample a broad range of configurational space~\cite{laio_escaping_2002,barducci_metadynamics_2011, shang_stochastic_2013,prinz_markov_2011}.
By applying additional potential on the region already sampled, it forces a system to explore over energy barriers~\cite{laio_escaping_2002,barducci_metadynamics_2011}.
They efficiently estimate the probabilities of sampled local minima or observable properties of an ensemble.
However, they cannot provide the probability estimates of multiple transition pathways and associated kinetic information.

To overcome the limitations of single-end methods, various double-end, or fixed-boundary-value, sampling approaches have been developed~\cite{berkowitz1983diffusion,pratt_statistical_1986,elber1987method,Czerminski1990,gillilan1992shadowing,cho1994construction,Olender1996,jonsson_nudged_1998,passerone_action-derived_2001,straub2002long,weinan2005finite,wales2002discrete,lee_kinetic_2003,faccioli_dominant_2006,Faccioli2008,vanden2010transition,avazquez_hcn_2015,a2015variational,lee_direct_2020,lee_finding_2017}.
The fixed-boundary-value formulation guarantees to find virtual trajectories connecting two end states if a calculation converges successfully, while the initial-value formulation does not. 
Berkowitz et al. suggested variational formula for the system with frictional resistance~\cite{berkowitz1983diffusion}, 
%Pratt suggested Metropolis paths identify transition state region~\cite{pratt_statistical_1986}.
Elber and Karplus suggested the Gaussian chain approach that minimizes the line integration of a potential along a trajectory~\cite{elber1987method}.
Czerminski and Elber added a repulsion constraint to the Gaussian chain approach to prevent replicas from aggregation~\cite{Czerminski1990}. 
%Cho et al. suggested the use of  where it 
To accelerate the convergence, the nudged elastic band (NEB) method was suggested~\cite{jonsson_nudged_1998}.
NEB applies an artificial force to the tangential component of a trajectory chain to remove the corner-cutting problem.
These methods find the minimum energy pathway (MEP) with the lowest energy barrier along the transition trajectory.  
However, the concept of a minimum energy pathway is not well-defined when multiple energy barriers exist between end states on a highly rugged energy landscape.
%These methods relied on the classical minimum action principle, which is an extremal principle. 
%In other words, the principle suggests physically accessible pathways correspond to the stationary points of classical action. 
%Therefore, the classical action can be maximized or minimized, which introduces computational ambiguity~\cite{passerone_action-derived_2001}. 

%Zimmerman suggested the concept of `connectivity graph' which identifies the intermediates as reachable by bond-breaking or bond-forming~\cite{zimmerman_automated_2013}. 
%Martinez-Nundez developed the `normal mode sampling method', where the transition state (TS) was searched by extrapolating the vibrational modes~\cite{avazquez_hcn_2015}. 
%Many of the methods gave artificial images leading the initial state to the final, physical interpretation of this pathway was lacking on those methods. 

Contrary to MEP-based approaches, pathway sampling methods based on the least action principles were proposed~\cite{berkowitz1983diffusion,gillilan1992shadowing,cho1994construction,Olender1996,passerone_action-derived_2001,eastman2001simulation,lee_kinetic_2003,faccioli_dominant_2006,Faccioli2008,fujisaki2010onsager,a2015variational,lee_direct_2020,lee_finding_2017}.
The variational Verlet algorithm~\cite{gillilan1992shadowing} and the Fourier component relaxation method that minimize the classical action of the trajectory expressed as Fourier (sine) components~\cite{cho1994construction} were suggested.
Olender and Elber suggested using Onsager Machlup action to investigate the long-time molecular trajectory~\cite{Olender1996}. 

Passerone and Parrinello~\cite{passerone_action-derived_2001} developed an action-derived molecular dynamics (ADMD) method, which finds low-action pathways via local optimization of a modified classical action with an energy conservation restraint term to keep the system's total energy constant throughout a reaction.
Lee and coworker~\cite{lee_kinetic_2003} developed a kinetic controlled ADMD method  introducing a restraining term to the modified action that satisfies the equipartition theorem.
However, these classical-action-based methods have an inherent limitation: the classical principle of least action is an extremal, not a minimum, principle. 
In other words, the classical action can be either minimized or maximized, which makes its computational outcome ambiguous. 
% Additionally, in a high temperature simulation, Brownian motions are naturally appear, making computationally difficult to simulate the long-time reaction. 

As a way to overcome this limitation of classical-action-based methods, pathway sampling approaches based on the Onsager-Machlup (OM) action~\cite{machlup_fluctuations_1953a,machlup_fluctuations_1953b, miller2007sampling} have been suggested.
the probability $p(\mathbf{x}_{B} \| \mathbf{x}_{A};t)$ for a state $\mathbf{x}_{A}$ to become another state $\mathbf{x}_{B}$ within a given time interval $ t $, can be described as follows~\cite{faccioli_dominant_2006,machlup_fluctuations_1953a, machlup_fluctuations_1953b,miller2007sampling}:
\begin{equation}
  p(\mathbf{x}_{B} \| \mathbf{x}_{A};t) = \int_{\mathbf{x}_A (\tau = 0)}^{\mathbf{x}_B (\tau = t)}{\mathcal{D}\mathbf{x}(\tau)e^{-i\hbar S_{\mathrm{OM}}\left(\mathbf{x}\left(\tau \right)\right)}}
  \label{eq:s_om}    
\end{equation}
where $S_{\mathrm{OM}} $ is the OM action and $\mathcal{D}$ is the notation for integrating all the possible pathways connecting two end states.
The probability of a transition from the state $\mathbf{x}_{A}$ to $\mathbf{x}_{B}$ is calculated by summing up the probability of all pathways $\sum_{s} e^{-S_\mathrm{OM}\left(\mathbf{x}\left( \tau \right)\right)/k_\mathrm{B}T}$~\cite{machlup_fluctuations_1953a,machlup_fluctuations_1953b,Bach1978,wio2013path}. 
For most cases, a state transition usually incorporates multiple pathways, such as multiple protein folding pathways.
The most dominant pathway is determined by the path with the lowest OM action, corresponding to the maximum probability of a transition~\cite{Bach1978,eastman2001simulation,faccioli_dominant_2006,miller2007sampling,adib2008stochastic,Faccioli2008,lee_finding_2017, lee_direct_2020}.
Lee and coworkers utilized an efficient global optimization method to identify the most dominant transition pathway corresponding to the global minimum of OM action and sample multiple less prevalent pathways~\cite{lee_finding_2017}.

A major bottleneck of the existing pathway search methods is their computational cost. %that they are computationally expensive. 
Calculating the action of a pathway requires the energy and gradient of every replica of a pathway. 
A small step size between replicas is necessary to generate smooth pathways.
Thus, there is a trade-off between trajectories' accuracy and computational efficiency.

%Recent Machine learning approach Due to the both algorithm and recent technological improvement on the computational machine learning field, atomic system based on machine learning are bursting out from the field. 

%Recent Gaussian approaches  Laude and Richardson et al. suggested similar in method yet, purpose is to calculate the Hydrogen hopping. In the study of Hydrogen moldecule splitting method~\cite{laude_ab_2018}.

To alleviate this computational cost issue, J\'{o}nsson and coworkers~\cite{koistinen_nudged_2017, garrido_torres_low-scaling_2019} suggested the one-image evaluation (OIE) method by combining the Gaussian process (GP) ~\cite{bernardo_bayesian_1999, ohagan_curve_1978} with NEB~\cite{jonsson_nudged_1998}. 
The OIE method finds the MEP by building a Gaussian potential energy surface (PES) approximating the actual PES iteratively.
They reduced the number of function calls  reduced by two orders of magnitude on an artificial 2D PES~\cite{garrido_torres_low-scaling_2019}.

Another major limitation of existing pathway search methods, such as NEB, is that they are \emph{local} optimization methods whose final results heavily depend on initial guesses on pathways.
Generally, conventional pathway search methods start from the linear interpolation between end states and perform local optimization of the initial pathway.
As a result, these approaches can find the correct minimum energy or action pathways only when a given PES is simple and smooth.
However, many complex systems have highly rugged PESs, which do not guarantee to locate the most dominant pathway near the linear interpolation between two states.
An extensive search on a pathway space is essential to avoid this problem.

To overcome these limitations of the pathway search methods, we introduce a novel transition-pathway-sampling method, Gaussian process action optimization (GPAO), by combining the GP algorithm with a global action optimization approach. 
Inspired by the OIE method, we only calculate the most uncertain images measured by the variance obtained with GP. 
Then the energy and forces of the images are used to construct the Gaussian PES.
Afterward, the intermediate images' potential energies, forces, and Hessian are estimated using an updated Gaussian PES. 
Pathways are optimized to have the minimum action using the predicted potential, forces, and Hessian.
Our method conducts direct OM action optimization with a total-energy-conservation restraint without using the original system's Hessian. 
%Gaussian processed action optimization (GPAO) samples multiple sub-optimal pathways, local minima of action, because it is based on an efficient global action method, the Action conformation space annealing (Action-CSA) algorithm~\cite{lee_finding_2017}.

The most significant advantage of GPAO is a dramatic reduction of the number of force calls by \emph{four-to-five orders of magnitude} than the conventional action optimization method while preserving their accuracy. % (Table~\ref{table:performance}).
%Owing to the efficiency of this routine, we can show that GPAO reduces the number of 1,200,000 function calls of a conventional action optimization method down to 12 to find a converged low-action pathway with comparable accuracy.
%GPAO reduces the number of 1,200,000 function calls of a conventional action optimization method down to 12 to find a converged low-action pathway with comparable accuracy.
%For example, the numbers of force calls to find $\mathcal{C}_\Theta^\mathrm{cls}$, $\mathcal{C}_S^\mathrm{OM}$, and $\mathcal{C}_\Theta^\mathrm{OM}$ using the conventional action optimization methods are 1,220,400, 586,200, and 993,900, while those of our GPAO method are only 12, 14, and 12, respectively. 
%Moreover, GPAO has advantage on not reqiring Hessian, iterative parameter searching is possible.  Second advantage of use is not requiring hessian information. @@Hessian increase@@  In other words, our GPAO method reduces the computational costs  of finding transition pathways by five orders of magnitude.
The significant reduction of the computational cost by GPAO allows us to find multiple transition pathways of a molecule---here, we choose alanine dipeptide---with a quantum-mechanical (QM) potential.
Additionally, we investigated the isomerization pathways of small organic molecules and the rearrangement pathways of atoms on a metallic surface. 
%For alaine dipeptide, a single force calculation of alanine dipeptide with \textsc{VASP} requires around 1741.4 CPU seconds. 
%Using GPAO, we can find the optimal transition pathway of alanine dipeptide within a single day. 
%Using DAO, however, it is estimated to took a year to complete the same calculation by the conventional action optimization method and the QM potential.
Throughout these examples, it is clearly identified that GPAO successfully identifies multiple physically accessible trajectories between the two end-states using QM potential energy.
To the best of our knowledge, this work is the first report, which identifies multiple conformational transition pathways of alanine dipeptide with an \textit{ab initio} potential.

% \section{Results and discussion}\label{sec:Method}
\section{Transition pathways on the M\"uller-Brown potential}

The M\"uller-Brown (MB) potential is a 2-dimensional model potential used to benchmark double-end methods\cite{muller_location_1979}.
%The 2D nature of MB is ideal for visualizing complex reaction mechanisms in a simple 2D map form. % Juyong: Not necessary.
To benchmark the efficiency of GPAO, we performed three calculations using GPAO, direct action optimization (DAO) without using a Gaussian potential, and the NEB method on the MB potential and compared their efficiency and accuracy.
The number of intermediate images, range of hyperparameters, and Gaussian kernel are set the same as previous studies~\cite{passerone_action-derived_2001, garrido_torres_low-scaling_2019}.
The MB potential is defined as follows:
\begin{equation}
  V\left(x,y\right) = \sum_{\mu=1}^{4}{A_\mu \exp[a_\mu \left(x-x_\mu^0\right)^2 + b_\mu \left(x-x_\mu^0\right) \left(y-y_\mu^0\right) + c_\mu\left(y-y_\mu^0\right)^2]}, 
\end{equation}
where $x_\mu^0=\left(1,\ 0,\ -0.5,\ -1\right)$, 
$y_\mu^0=\left(0,\ 0.5,\ 1.5,\ 1\right)$, 
$a_\mu=\left(-1,\ -1,\ -6.5,\ 0.7\right)$, 
$b_\mu=\left(0,\ 0,\ 11,\ 0.6\right)$, 
$c_\mu=\left(-10,\ -10,\ -6.5,\ 0.7\right)$, and $A_\mu=\left(-2,-1,-1.7,\ 0.15\right)$.

\begin{figure*}[htb!]
  \begin{adjustbox}{width=1\textwidth}
    \begin{tikzpicture}

      \node[anchor=south west] (X) at (0,0){
        \includegraphics[trim=70 45 80 40, clip, width=1\textwidth]{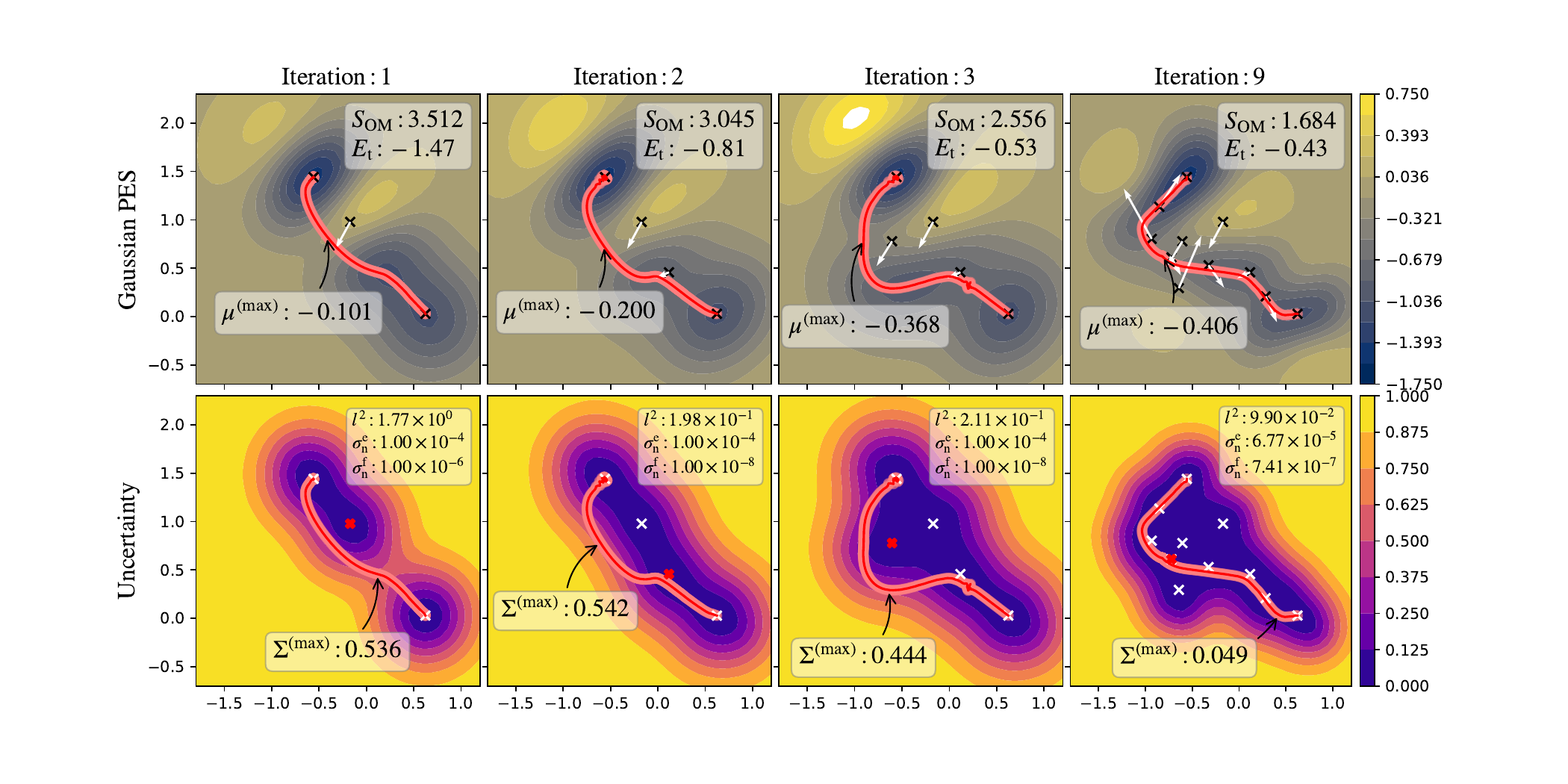}
      };
      \begin{scope}[x={(X.south east)},y={(X.north west)}]%
        \node[anchor=north west] (graph) at (0,0){
          \includegraphics[trim=0 0 0 5, clip, width=0.99\textwidth]{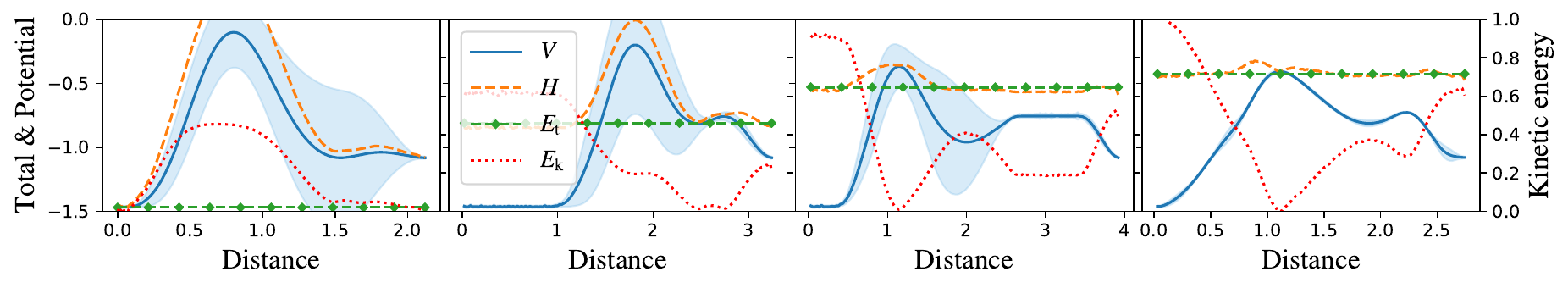}};
        
        \node[text width=0in] (A) at (0.01,0.9) {{(a)}};
        \node[text width=0in] (B) at (0.01,0.4) {{(b)}};
        \node[text width=0in] (C) at (0.01,0.0) {{(c)}};
        
      \end{scope}%
    \end{tikzpicture}
  \end{adjustbox}
  \caption{The demonstration of finding a transition pathway using GPAO.
    The expectation value of the (a) Gaussian PES map, (b) uncertainty map, and (c) energy graph along the pathway are plotted from top to bottom. 
    The map and graph at iterations 1, 2, 3, and 9 are plotted from left to right. 
    The red line at the Gaussian PES indicates trajectory, the black crosses are positions where data is acquired, and the white arrows are the force data of the corresponding coordinates. 
    On the uncertainty map, we marked additional red crosses corresponding to the latest position where data are acquired in addition to the white crosses marked at identical locations on all Gaussian PESs.
    In the energy graph, the blue line shows potential energy, light blue means 95\% confidence range of potential energy, the red dot indicates kinetic energy, the orange dashed line is the total energy and the green diamond line expresses target energy.
  }
  \label{fig:MBGPAO}
\end{figure*}
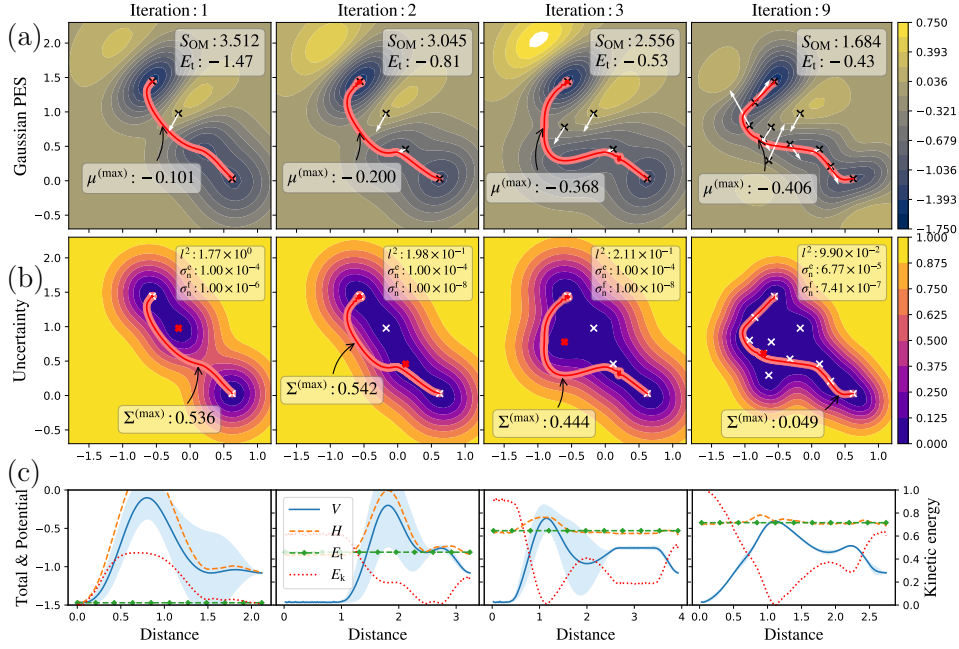

% ############### Process MB
The process of how GPAO minimizes the MEP's uncertainty and approximates the MB potential is illustrated in Figure~\ref{fig:MBGPAO}.
At iteration 1, the pathway is optimized with the three data points: the initial, the final, and a random intermediate state between the two. 
Due to the lack of data points, potential energy values estimated along the trajectory are highly uncertain. 
The uncertainty range (light blue intervals) of the estimation at iteration 1 is much wider than iteration 9 (Figure~\ref{fig:MBGPAO}b).
The main idea is to calculate the energy and forces of the point that minimizes the uncertainty of a trajectory most. 
Thus, at iteration 2, GPAO collects data at the most uncertain region of the previous iteration (marked with a red cross in Figure~\ref{fig:MBGPAO}b).
After the update, the OM action of a pathway is minimized on the updated Gaussian PES, whose hyperparameters are optimized with new data points. 
After hyperparameter optimization, at iteration 3, the energy and forces of the most uncertain points in the pathway are added to the database (red crosses in Figure~\ref{fig:MBGPAO}b), and the process is iterated. % at iteration 2).
At iteration 9, the pathway is considered as converged since the uncertainty criteria, $\mathbf{\Sigma}^{\mathrm{(max)}}<0.05$, are satisfied.
In total, GPAO requires 11 force calls.

%Since GPAO is an iterative method, a subset of its hyperparameters, i.e., the targeted total energy of the system, $E_\mathrm{t}$, is adjusted by following equation,
The targeted total energy of the system, $E_\mathrm{t}$, one of the hyperparameters of the method, is adjusted during iterations based on the following equation:
$$ E_\mathrm{t}^{(i)} = \frac{E_\mathrm{t}^{(i - 1)} + \boldsymbol{\mu}^{\mathrm{(max)}}}{2},$$
where $E_\mathrm{t}^{(i)}$ is the target energy of $i$th iteration, $E_\mathrm{t}^{(i-1)}$ is the target energy of iteration $i-1$, and $\boldsymbol{\mu}^{\mathrm{(max)}}$ is the estimated maximum potential energy of iteration $i-1$.
%Figure~\ref{fig:MBGPAO}(c) shows how GPAO approximates $E_\mathrm{t}$ to $V^{\mathrm{(max)}}$ iteratively (Eq. XX). %% approximates A to B?
Figure~\ref{fig:MBGPAO}(c) shows how GPAO approximates $E_\mathrm{t}$, iteratively. %% approximates A to B? 
At iteration 1, $E_\mathrm{t}$ is set to $-1.47$, the minimum energy between the initial and the final image. 
GPAO sets $\boldsymbol{\mu}^{\mathrm{(max)}} = -0.10$ from the trajectory optimized with three data points. 
Using this information, at iteration 2, $E_\mathrm{t}$ is raised to $-0.81$, which is the half of $E_\mathrm{t} + \boldsymbol{\mu}^{\mathrm{(max)}}$ of iteration 1.
Thus, GPAO estimates $\boldsymbol{\mu}^{\mathrm{(max)}}$ to be $-0.20$ with the four data points. 
Similarly, at iteration 3, $E_\mathrm{t}$ is set to the half of $E_\mathrm{t} + \boldsymbol{\mu}^{\mathrm{(max)}}$ of iteration 2, $-0.368$. 
%The maximum uncertainty at iteration 1 ($\mathbf{\Sigma}^{\mathrm{(max)}}=5.15\times 10^{-1}$) is smaller than that at iteration 9 ($\mathbf{\Sigma}^{\mathrm{(max)}}=9.47\times10^{-2}$) (Figure~\ref{fig:MBGPAO}b).

\begin{figure}[htb!]
  \includegraphics[trim=0 20 30 40, clip, width=0.5\textwidth]{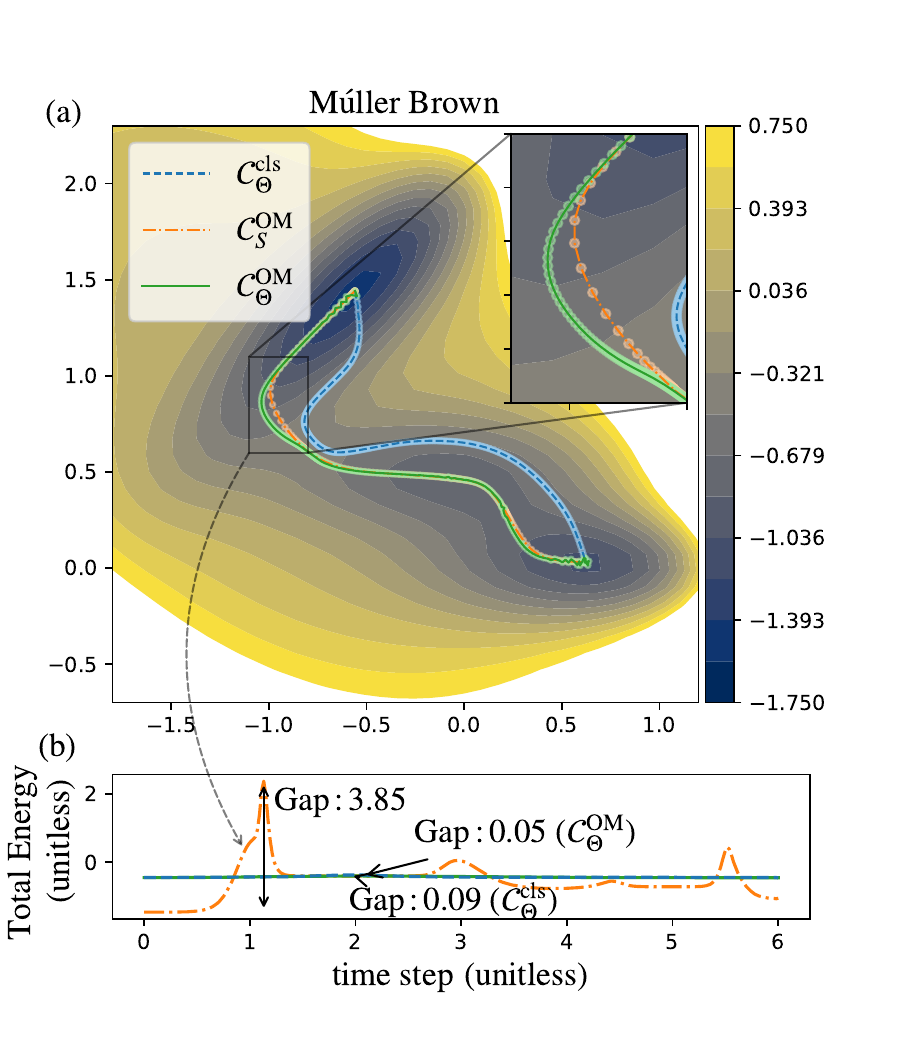}
  \caption{\label{fig:SomSclsComparison} 
    A comparison of minimum action pathways obtained with different actions. 
    Pathways obtained with classical action with energy conservation restraint ($S_\mathrm{cls}$, blue dashed line), OM action ($S_\mathrm{OM}$, orange dash-dotted line), and OM action with energy conservation constraint ($\Theta_\mathrm{OM}$, solid green line). 
    (a) The minimum action pathways between two stable potential energy minima on the M\"uller-Brown potential and a zoomed image near the saddle point.
    (b) The energy profile of the three trajectories obtained with three different actions: classical action, OM action, and the modified OM action with total energy conservation term.}
\end{figure}

Using the target energy obtained from GPAO, we also obtain pathways using DAO with three actions: classical action with total energy conservation restraint $S^{\mathrm{cls}}$ (blue dashed line in Figure~\ref{fig:SomSclsComparison}),  OM action $S^{\mathrm{OM}}$ (orange dash-dotted line in Figure~\ref{fig:SomSclsComparison}), 
and OM action with total energy conservation restraint ${\Theta}^{\mathrm{OM}}$ (solid green line in Figure~\ref{fig:SomSclsComparison}). 
The trajectories obtained with direct optimization of three actions are denoted as $\mathcal{C}_\Theta^\mathrm{cls}$, $\mathcal{C}_S^\mathrm{OM}$, and $\mathcal{C}_\Theta^\mathrm{OM}$, and the trajectories obtained with GPAO are denoted as GP-$\mathcal{C}_\Theta^\mathrm{cls}$, GP-$\mathcal{C}_S^\mathrm{OM}$, and GP-$\mathcal{C}_\Theta^\mathrm{OM}$. 

%################ Accuracy MB 
\begin{table}[htb!]
  \caption{
    A comparison of performance between the GP-assisted algorithm and the conventional action algorithm.
    The calculation is conducted on the  M\"uller-Brown potential.
    The number of force call is the number of actual function evaluations during the optimization process, $V^{\mathrm{(max)}}$ is the maximum potential energy of a pathway, $S_{\mathrm{OM}}$ is the OM action value of a pathway. 
    Distance to $\mathcal{C}_S^\mathrm{OM}$ is a Fr\'echet distance to $\mathcal{C}_S^\mathrm{OM}$ and distance to non-GP is a Fr\'echet distance from the pathway optimized via non-GP-assisted method}
  \label{table:performance}
  \begin{tabular}{cccccc}
    \hline
    & Number of force calls &  $V^{\mathrm{(max)}}$ & $S_{\mathrm{OM}}$ & Distance to $\mathcal{C}_S^\mathrm{OM}$ & Distance to non-GP \\
    \hline

    $\mathcal{C}_\Theta^\mathrm{cls}$          & $300\times 4068$ & $-0.400$  & 14.49 & 0.250  & -  \\
    $\mathcal{C}_S^\mathrm{OM}$                & $300\times 1954$ & $-0.406$  & 1.387 &     -  & -  \\
    $\mathcal{C}_\Theta^\mathrm{OM}$           & $300\times 3313$ & $-0.406$  & 1.652 & 0.067  & -  \\ \hline
    GP - $\mathcal{C}_\Theta^\mathrm{cls}$     & $12$             & $-0.401$  & 11.08 & 0.250  & 0.006 \\
    GP - $\mathcal{C}_S^\mathrm{OM}$           & $14$             & $-0.405$  & 1.187 & 0.020  & 0.082 \\
    GP - $\mathcal{C}_\Theta^\mathrm{OM}$      & $12$             & $-0.406$  & 1.684 & 0.083  & 0.042 \\ \hline
    $\mathcal{C}^\mathrm{NEB}$                 & $20 \times 30$   & $-0.407$  &    -  &   -    &  -  \\
    GP-$\mathcal{C}^\mathrm{NEB}$              & $12$             & $-0.412$  &    -  &   -    & 0.0766 \\
    \hline
  \end{tabular}
\end{table}

The most remarkable result is that GPAO requires significantly fewer force calls than DAO (Table~\ref{table:performance}).
GP-$\mathcal{C}_\Theta^\mathrm{cls}$ requires 12 force calls, while $\mathcal{C}_\Theta^\mathrm{cls}$ requires around 1,200,000 force calls.
Similarly, GP-$\mathcal{C}_S^\mathrm{OM}$ and GP-$\mathcal{C}_\Theta^\mathrm{OM}$ require 14 and 12 force calls respectively, while $\mathcal{C}_\Theta^\mathrm{cls}$ and $\mathcal{C}_\Theta^\mathrm{cls}$ require 600,000 and 900,000 force calls respectively.
Overall, GPAO reduced the number of \emph{force calls in four to five orders of magnitude}.

The GPAO results are highly similar to those obtained with DAO with the significantly reduced numbers of force calls.
Overall, the Fr\'etchet distances between the pathways obtained with GPAO and DAO are less than 0.08\% of the pathways' total distances.
The distance between $\mathcal{C}_\Theta^\mathrm{cls}$ and GP-$\mathcal{C}_\Theta^\mathrm{cls}$ is 0.006.
Similarly, the distances between $\mathcal{C}_S^\mathrm{OM}$ and $\mathcal{C}_\Theta^\mathrm{OM}$ and their counterpart obtained with GPAO are 0.082 and 0.042, respectively.
GP-$\mathcal{C}_\Theta^\mathrm{cls}$ deviates from $\mathcal{C}_\Theta^\mathrm{cls}$ less than a Fr\'echet distance of 0.01.
All four GP-assisted pathways show a deviation of less than a Fr\'echet distance of 0.1 than the DAO results.
These results clearly demonstrate that approximating PES with GP is accurate enough to locate the transition pathways with low action on a given PES.

%The most remarkable result is a difference in a number of force calls required between GPAO and DAO.

The convergence of GPAO depends on the number of data points near the MEP, which should be large enough to approximate the PES accurately.
Overall, GPAO requires similar numbers of force calls regardless of the selection of the objective function (Table~\ref{table:performance}). % NEB is not an action-based method.
For example, the results of the GP-assisted NEB method, GP-$\mathcal{C}^\mathrm{NEB}$, showed similar numbers of force calls to reproduce the result of the conventional action optimization method. 
Indeed, the number of data points required to find an accurate MEP on the MB potential is also between 10 and 20. 
These results imply that the type of local relaxation method does not significantly affect the efficiency of GPAO. 
In contrast to GPAO, the calculation cost of DAO is strictly proportional to the number of images in a pathway.

After confirming that GPAO successfully optimizes a given action of a pathway, we compare the characteristic of $\mathcal{C}_\Theta^\mathrm{cls}$, $\mathcal{C}_S^\mathrm{OM}$, and $\mathcal{C}_\Theta^\mathrm{OM}$ (Figure~\ref{fig:SomSclsComparison}a). 
%optimized pathways obtained with three actions: $S^\mathrm{OM}$, $S^\mathrm{OM}$, and $\Theta^\mathrm{OM}$ (Figure~\ref{fig:SomSclsComparison}(a)). %optimization results of three actions: the modified classical action and the OM action with and without the total energy conservation term. 
The results demonstrate that $\Theta_\mathrm{OM}$ successfully conserves the system's total energy throughout the pathway (Figure~\ref{fig:SomSclsComparison}b). 
The magnitude of total energy fluctuations of $\mathcal{C}_\Theta^\mathrm{OM}$ is reduced by 98.7\% than that of $\mathcal{C}_S^\mathrm{OM}$. 
The action value of $\mathcal{C}_\Theta^\mathrm{OM}$ increases only about 19\%, and the Fr\'echet distance to $\mathcal{C}_S^\mathrm{OM}$ is less than 0.067 (Table~\ref{table:performance}).
Also, 95\% of the intermediate images of $\mathcal{C}_\Theta^\mathrm{OM}$ are located within a distance of 0.013 from $\mathcal{C}_S^\mathrm{OM}$. 
This result demonstrates that the total energy conservation term affects little to the shape of the pathway while conserving the total energy of the system.
The results indicate that enforcing the energy conservation restraint on $\mathcal{C}_S^\mathrm{OM}$ gives only little effect on finding low $V^\mathrm{(max)}$ and $S_{\mathrm{OM}}$. 
%The Fr\'echet distance between two trajectories is only 0.067. 

Figure~\ref{fig:SomSclsComparison}b shows the total energy profiles of three trajectories and their energy gaps, the energy differences between the maximum and the minimum total energies of pathways.
The energy gaps of $\mathcal{C}_\Theta^\mathrm{cls}$, $\mathcal{C}_S^\mathrm{OM}$ and $\mathcal{C}_\Theta^\mathrm{OM}$ are 0.09, 3.85 and 0.05, respectively. 
The large spikes of the total energy are observed between the time steps 0.8 $\sim$ 1.1, where the trajectory climbs up the energy barriers of the PES.
The main reason for this huge spike of the total energy is the second term of the $S_\mathrm{OM}$ equation (Eq.~(\ref{eq:Som})). 
The first term in Eq.~(\ref{eq:Som}) prefers to minimize atoms' forces, and the third term keeps the velocities between images low.
The second term in Eq.~(\ref{eq:Som}) favors following a concave potential energy surface along its ongoing direction. 
More specifically, if a pathway has to climb up a PES during a transition, it prefers to climb up the surface fast to minimize action. 
This tendency is clearly manifested in Figure~\ref{fig:SomSclsComparison}a. 
Right before $\mathcal{C}_S^\mathrm{OM}$ reaches the saddle point with the maximum potential energy of the PES, the images are largely separated because of the large velocities of corresponding images.
In contrast, the images of $\mathcal{C}_\Theta^\mathrm{OM}$ near the saddle point are almost continuous and uniformly distributed. 
Uniform distances between images prevent large fluctuations of total energy.
In other words, GPAO avoids the occurrence of unphysical energy spikes observed with $\mathcal{C}_S^\mathrm{OM}$, but still maintains $S^\mathrm{OM}$ low.
The action values of $\mathcal{C}_\Theta^\mathrm{cls}$, $\mathcal{C}_S^\mathrm{OM}$ and $\mathcal{C}_\Theta^\mathrm{OM}$ are 14.49, 1.387, and 1.652, respectively. 
The potential energies at the saddle point, $V^\mathrm{(max)}$, of $\mathcal{C}_\Theta^\mathrm{cls}$, $\mathcal{C}_S^\mathrm{OM}$, and $\mathcal{C}_\Theta^\mathrm{OM}$ are $-0.400$, $-0.406$, and $-0.406$, respectively (Table~\ref{table:performance}).
Our results demonstrate that unphysical increases of atomic velocities to lower the second term can be suppressed effectively with the total energy conservation restraint without changing trajectories. 
%Furthermore, Figure~\ref{fig:SomSclsComparison}a shows how two pathways of $\mathcal{C}_S^\mathrm{OM}$ and $\mathcal{C}_\Theta^\mathrm{OM}$ overlap each other.
Furthermore, two pathways, $\mathcal{C}_S^\mathrm{OM}$ and $\mathcal{C}_\Theta^\mathrm{OM}$, overlap significantly each other.

\section{Isomerization pathways of small organic molecules}
\begin{figure*}[htb!]
   \begin{adjustbox}{width=0.9\textwidth}
     \includegraphics{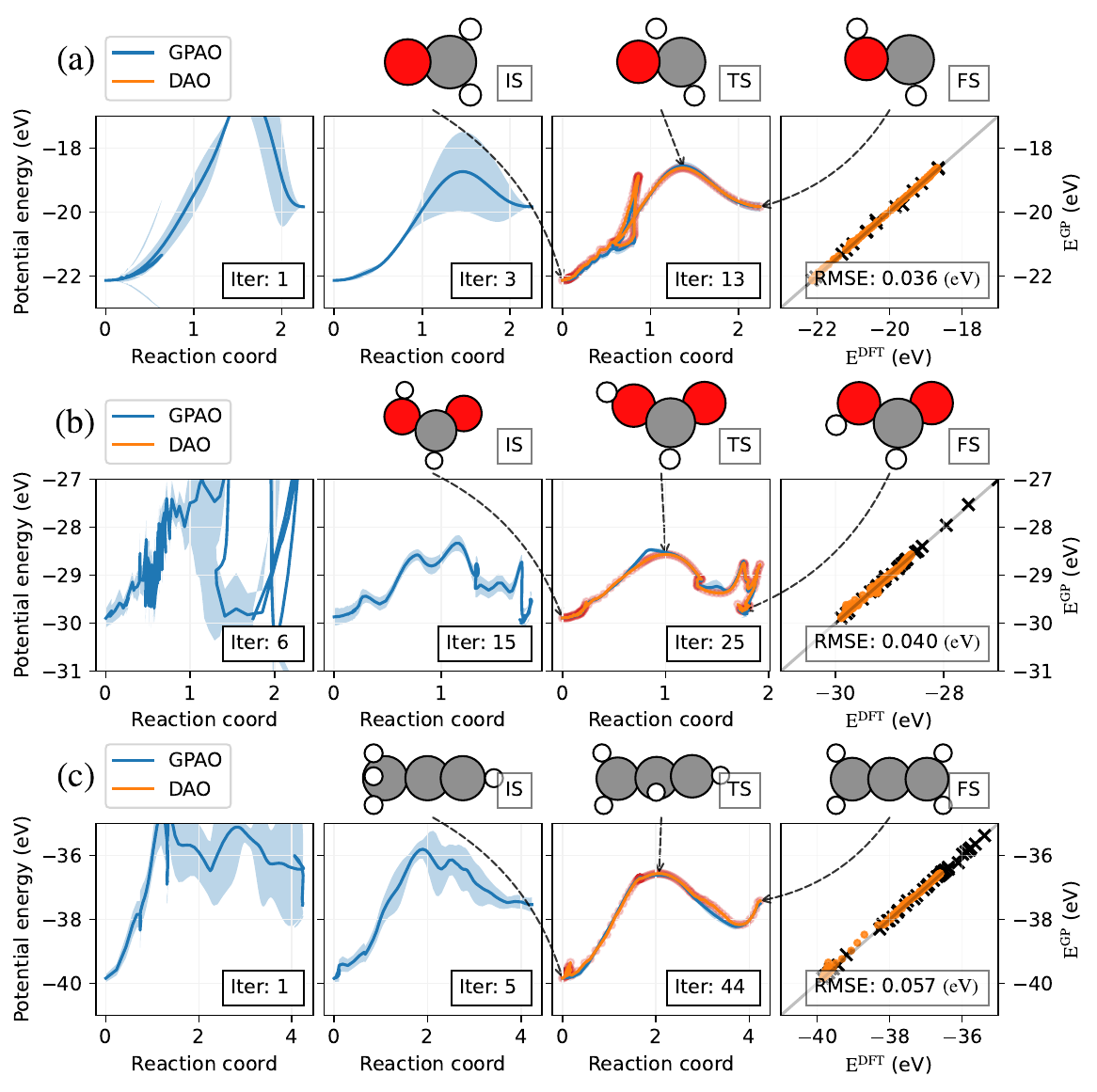}
   \end{adjustbox}
%  \begin{adjustbox}{width=1\textwidth/2}
%    \includegraphics{Isomerization2/FormicAcid_E.pdf}
%  \end{adjustbox}\llap{\parbox[b]{1\textwidth/2}{(b)\\\rule{0ex}{2.0in}}}
%
%  \begin{adjustbox}{width=1\textwidth/2}
%    \includegraphics{Isomerization2/Propyne_E.pdf}
%  \end{adjustbox}\llap{\parbox[b]{1\textwidth/2}{(c)\\\rule{0ex}{2.0in}}}
  
  %\begin{adjustbox}{width=0.5\textwidth}
  %\includegraphics[width=1\textwidth]{Isomerization/AuGPAO_E.pdf}
  %\end{adjustbox}
  \caption{A comparison of minimum action pathway calculation results obtained with GPAO and DAO on the isomerization reactions of (a) formaldehyde, (b) formic acid, and (c) propyne.
  For each molecule, potential energy profiles of GPAO at three different iterations are plotted from left to right. 
  At third, both the results obtained with GPAO and DAO are drawn together with three arrows pointing to each reaction coordinates of their atomic configurations, initial state (IS), the transition state (TS), and the final state (FS) of isomerization reaction.
  On the rightmost graph, a comparison between DFT data and its prediction by Gaussian PES constructed with GPAO is drawn with RMSE. 
  DFT data used for comparison is gathered from the trajectory obtained from DAO. 
  Black crosses are the data gathered by GPAO. 
  %Blue solid line is from GPAO's trajectory, orange solid line is from DAO's trajectory, black cross indicate the data point gathered by GPAO, and orange dots indicates the DFT data on the DAO`s trajectory. 
  Hydrogen, carbon, and oxygen are colored white, grey, and red.
  }
  \label{fig:organic_E}
\end{figure*}

% #### Intro MOL
To further validate the advantages of GPAO in application to real molecular systems, we benchmark the computational cost and characteristics of pathways obtained with DAO and GPAO using density functional theory (DFT) on the isomerization of small molecules: formaldehyde, formic acid, and propyne (Figure~\ref{fig:organic_E}). 
We perform GPAO first, and the resulting target total energy is used for DAO calculations.
Additionally, the final trajectory obtained from a GPAO calculation is used as the initial trajectory for DAO to reduce computational cost. 
In this way, the difference between trajectories approximated with GPAO and true ground state obtained with DAO is compared. 

%Also, it gives a qualitative computational benefit compared to GPAO. %% Don't know what does this mean...
% or lower bound of the computational cost of DAO. 

% Figure explanation
Atomic configuration of initial states (IS), transition state (TS), and final state (FS) of each molecule are drawn in  Figure~\ref{fig:organic_E} with the arrows pointing to corresponding reaction coordinates.   
GPAO begins with the five data, two initial and final states and three points between them.
During the process, the potential profile of the trajectory quickly converges as the data accumulate throughout iterations.
In Figure.~\ref{fig:organic_E}, the potential energy profiles of GPAO (solid blue line) and their variance (light blue shade) slowly converge to those of DAOs (solid orange line).
In the rightmost graph in Figure~\ref{fig:organic_E}, Gaussian PES constructed with this process are compared with true DFT data sampled along the DAOs trajectory (orange dot). 
In all three cases, using only, comparably a few data points sampled with the GPAOs process (black cross), root mean square error (RMSE) is less than $0.06$ eV, or $0.01$ eV/atom. 

% #### Results Mol
A comparison of the number of force calls, energy barrier, and actions obtained with GPAO and DAO is summarized in Table~\ref{table:organic_examples}. 
The number of force calls for DAO of formaldehyde, formic acid, propyne are $3.52 \times 10^5$, $3.70 \times 10^5$, and $4.73 \times 10^5$, respectively.
Meanwhile, the numbers of force calls conducting GPAO counterpart are $18$, $30$, and $49$, respectively. 
Thus, the overall computational cost of GPAO is less than that of DAO by \emph{two to three orders of magnitude}, while the error of OM action and energy barrier is suppressed under 5\%. 
All the example shows consistent improvement in computational efficiency with comparable accuracy of GPAO.

% ####### Hessian Mol
One of the major challenges in performing DAO with DFT is an enormous computational cost for Hessian calculations.
The Hessian matrix can be obtained analytically for simple potentials such as the MB potential. %the extra calculation cost to get Hessian is ignored by simply applying the analytical derivatives.
%Unfortunately, in the DFT method, there is no good way of getting Hessian without additional calculation.
However, obtaining a Hessian matrix is not straightforward for a DFT potential in general. %without additional calculation.
%One way to safely calculate the Hessian is to use the finite difference method.
Thus, a finite difference method is used in this study.
To obtain a Hessian matrix using the central difference method, $D$ additional force calls are needed, where $D$ is the degree of freedom.
%For example, the formaldehyde needs 9 additional force calculations for each Hessian calculation ((A total number of atoms $4$ $-$ one fixed atom $1$) $\times 3$).
For example, for the Hessian calculations of formic acid and propyne, 12 and 18 additional force calls are required.

% ### Fixation Mol
The Cartesian coordinates of atoms are used as the input for the Gaussian PES kernel.
%In GP, the Cartesian atomic position does not preserve rotation, translation symmetry.
To satisfy the rotational and translational invariance of the GP process, the position of an arbitrary atom of a system is fixed.
%This makes our coordinate relative to the fixed atom, thus removing the need for translation symmetry of the system.
When an atom is not fixed, GPAO keeps searching the same geometry but slightly translated position, making the convergence of GPAO difficult.

\begin{table}[htb!]
\caption{
  Accuracy and performance comparison of DAO and GPAO.
  The isomerization of organic molecules is calculated using VASP. %and metals.  %% WHAT DOES THIS MEAN?
  Note that * indicates that the initial trajectory is set to the results of the GPAO counterpart. 
}
\label{table:organic_examples}
\begin{tabular}{cccccc}
  \hline
  & Dim & Number of force calls & RMSE &  $\mathrm{Energy\ barrier}$ & $S_{\mathrm{OM}}$   \\
  \hline
  DAO-Formaldehyde   & 9 & *$150 \times 9 \times 261$   & -       & $3.51$ eV   & $22.04$ \\
  DAO-Formic acid    & 12 & *$150 \times 12 \times 220$ & -       & $1.32$ eV & $21.48$ \\
  DAO-Propyne        & 18 & *$150 \times 18 \times 175$ & -       & $3.27$ eV & $19.89$ \\ \hline
  GPAO-Formaldehyde  & 9 & $18$                         & 0.036 eV&  $3.57$ eV & $19.70$                         \\
  GPAO-Formic acid   & 12 & $30$                        & 0.040 eV & $1.42$ eV & $25.45$                        \\
  GPAO-Propyne       & 18 & $49$                        & 0.057 eV & $3.23$ eV & $19.75$                        \\
\hline
\end{tabular}
\end{table}

%In the dimensional, someone might peculiar about a number of force call conducted in 2D example is highly Comparison to 2D example, Reason for deviation is because Gaussian PES goes over fitting. Problem remains that how to cope with error parameters with overfitting issue is remain but the overall it looks very similar. 

\section{Surface reactions}
% ########### Intro  Surface 
In the third class of examples, the diffusion and rearrangement of atoms on a metallic surface are studied.
The potential energies of metallic systems are calculated using effective medium theory (EMT) in the Atomic Simulation Environment (ASE) package~\cite{larsen_atomic_2017}.
Using the EMT potential, we compare the characteristics of DAO and GPAO results on two examples: `Au hopping' and `Pt rearrangement'~\cite{garrido_torres_low-scaling_2019}.

\begin{figure*}[htb!]
    \begin{adjustbox}{width=\textwidth}
    \includegraphics{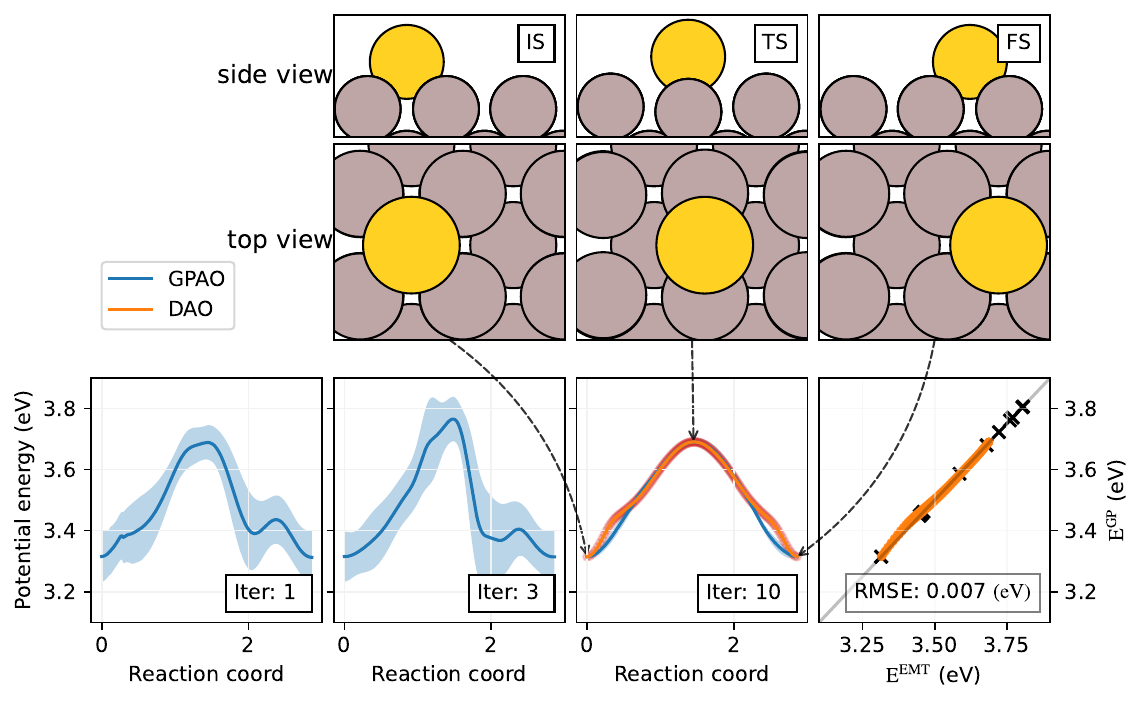}
    \end{adjustbox}
    %2.5in
  \caption{A comparison of low action pathways obtained with GPAO and DAO of a gold atom hopping on the aluminum fcc (100) surface.
  A side view and a top view of the surface reaction and its potential energy profiles are plotted. 
  The potential energy profile of GPAO (solid blue line) and its variance (light blue shade) at the iteration 1, 3, and 10 (Iter: 1, 3, and 10) are plotted from left to right.
  At iteration 10, DAOs potential energy profile (solid orange line) are drawn together with GPAOs one, with three additional arrows coming from the initial state (IS), transition state (TS), and final state (FS) to the corresponding reaction coordinate (Reaction coord).
  Finally, DFT data gained from GPAO (black cross) are compared with DFT data of DAOs trajectory (orange dot) at the rightmost graph. 
  The gold atom is colored in yellow, and aluminum is colored in brown.}
  \label{fig:gold}
\end{figure*}

% #######  System explain Gold
In the `Au hopping' example, a gold atom jumps to its neighboring site on the aluminum fcc (100) surface. 
To visualize the diffusive pathway, the atomic configurations at the key states are plotted in Figure~\ref{fig:gold}. 
From left to right, the atomic configurations of the initial state (IS), transition state (TS), and the final state (FS) watching from the top and the side are illustrated.
For both IS and FS, the gold atom is propped in the middle of four aluminum atoms. 
In TS, the gold atom is on the saddle point supported by two aluminum atoms. 
Two aluminum atoms right under the gold atom sink slightly, while the other two aluminum atoms, far from the gold atom, slightly protrude along the z-axis.
This shift lowers the energy barrier to 0.4 eV. 
The GPAO trajectory successfully reconstructs the shift of potential energy barrier, leading to an almost identical trajectory to the DAO trajectory.

% ### method & results & discussion Gold
Below the atomic configuration, energy profiles at iterations 1, 3, and 10 (Iter: 1, 3, and 10 in Figure~\ref{fig:gold}) are plotted.
While optimizing action, only the gold atom and its neighboring four aluminum atoms are allowed to move and the other atoms are fixed reducing the degree of freedom $D=15$. 
At iteration 1, GPAO optimizes a trajectory on a Gaussian PES, which is constructed with five initial data points. 
The initial data are IS, FS, and additional three random points between the two end states.
The energy profile at this stage highly deviates from that of DAO, and it has high uncertainty.
As the calculation proceeds, the GPAO trajectory and energy profile converge to those of DAO.
At iteration 10, the trajectory converges to the DAO trajectory with a covariance less than 0.05 eV. 
As a result, using only 13 data points, GPAO accurately predicted the saddle points.
%Using Gaussian PES constructed from these 13 points, we compare the EMT data acquired from DAO. 
The EMT data and predicted potential with GP of the images of the DAO trajectory are compared. 
For 15 images, the RMSE of potential prediction is 70 meV, or 5 (meV/atom). 
Compare to the common neural network potential, the number of data required to find the energy barrier is relatively small. 
This implies that our iterative searching scheme is highly effective for investigating surface hopping problems. 
%This particularly useful when the price for gathering each data points is expensive. 

\begin{figure*}[htb!]
\begin{adjustbox}{width=\textwidth}
    \includegraphics{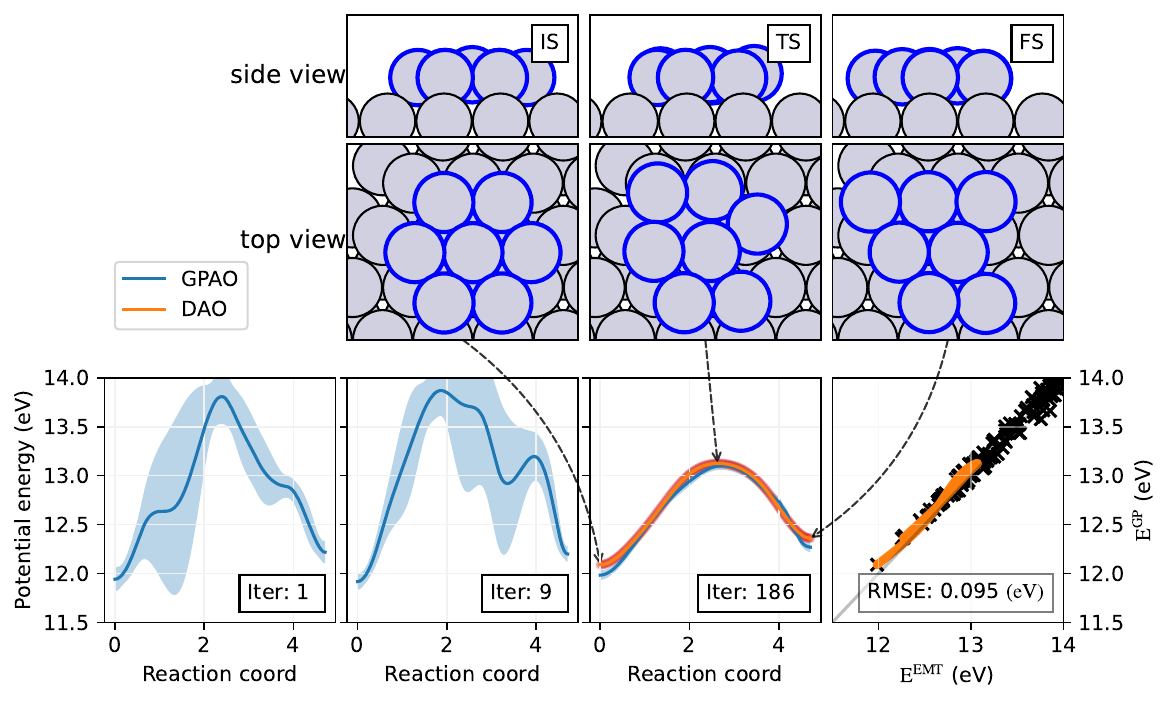}
\end{adjustbox}
%2.5in
  \caption{A comparison of rearrangement of platinum atoms on its surface.
  The potential energy profiles of GPAO (solid blue line) and their variances (light blue shade) at the iteration 1, 3, and 10 (Iter: 1, 9, and 186) are plotted.
  At iteration 186, DAO potential energy profile (solid orange line) is drawn together with that of GPAO, with three additional arrows indicating the initial state (IS), transition state (TS), and final state (FS) to the corresponding reaction coordinate (Reaction coord).
  Finally, predicted energies sampled from GPAO potential surface (black cross) are compared with the DFT of DAO trajectory (orange dot) at the rightmost graph. 
  The platinum atom is colored with grey and surface atoms are emphasized with blue edges.}
  \label{fig:hex}
\end{figure*}

%% ############# System explain platinum hex
In the `Pt rearrangement' example, seven platinum atoms packed as a hexagonal shape on the fcc (111) surface are rearranged to a parallelogram shape. 
Among 71 atoms, seven atoms on the surface are set to move freely, setting the degree of the freedom as $D=21$. 
To lower the calculation cost for the Gaussian kernel, surface platinum atoms under the seven hexagon atoms are fixed.
In Figure~\ref{fig:hex}, from left to right, IS, TS, and FS of `Pt rearrangements' are drawn emphasizing moving atoms with a blue edge. 
In IS and FS, seven platinum atoms on the surface are located at symmetry sites.
%For each platinum atom are resides above the threefold center,
In TS, the center atom is misplaced from its symmetric sites, while maintaining a two-fold symmetry.
This lowers the energy barrier to 1.0 eV.

%### Results
Below the atomic configurations in Figure~\ref{fig:hex}, the energy profiles of GPAO at iterations 1, 9, and 186 (Iter:1, 9, and 186) are illustrated.
At iteration 1, GPAO optimizes the trajectory from five data points: IS, FS, and three linearly interpolated points between the two.
The energy barrier at this stage is nearly 2 eV, and the confidence for this prediction is low as expected. 
As low energy configurations accumulate, the trajectory converges to the DAO trajectory. % by adding up the data points suspect to have low energy configuration. 
%This guessing process is often wrong, fail to search the low energy configuration. 
The black cross mark stacked at the RMSE graph is these attempts to correctly locate the configuration near-optimal pathways. 
At iteration 186, GPAO converges and successfully locates the MEP. 
A comparison between the predicted energies on the Gaussian PES and EMT data sampled from the DAO trajectory is plotted on the right side of Figure~\ref{fig:hex}. 
The RMSE between two calculations is 95 meV, or 5 meV/atom.
%This is another example of GPAO that creates a comparable trajectory to DAO.
These results show that GPAO predictions are in great agreement with the EMT results.

\begin{table}[htb!]
  \caption{Accuracy and performance comparison of DAO and GPAO.}
  \label{table:surface_comparison}
  \begin{tabular}{cccccc}
    \hline
    & Dim & Number of force calls &  RMSE & $\mathrm{Energy\ barrier}$ & $S_{\mathrm{OM}}$  \\
    \hline
    DAO-`Au hopping'   & 15 & $150 \times 15 \times 187$ & -& $0.376$ eV& $4.57$ \\
    DAO-`Pt rearrange' & 21 & $150 \times 21 \times 244$ & -& $1.039$ eV& $13.16$ \\ \hline
%    DAO-`Pt diffusion'& 48 & $150 \times 48 \times 194$ & $0.70$ eV & $23.62$ \\
    GPAO-`Au hopping'& 15   & $15$  & 0.007 eV  & $0.375$ eV & $4.72$ \\
    GPAO-`Pt rearrange'& 21 & $191$ & 0.095 eV  & $1.119$ eV & $15.09$ \\
%    GPAO-`Pt diffusion'& 48 & $120$ & $0.85$ eV & $24.79$ \\
    \hline
  \end{tabular}
\end{table}

% ######## Reuslts & Discussion again
%##### Table 1 explain Dimensional problem surface method & results & discussion 
To validate the advantage of the GPAO method, the results of both DAO and GPAO are summarized in Table~\ref{table:surface_comparison}.
The degree of freedom, a number of force calls, energy barriers, and the OM actions of final trajectories obtained with GPAO and DAO are tabulated.
The degrees of freedom of `Au hopping' and `Pt rearranges' are 15 and 21, respectively. 
%This affects the Hessian calculation critically, for each example, to get a Hessian of a point, it requires the additional $D$ number of force calls. 
Because DAO requires Hessian calculations, the numbers of force calls for DAO calculations of `Au hopping' and `Pt rearrange' are 420,750 and 768,600 respectively, while those of GPAO are 15 and 115.
GPAO predicts the energy barriers of three examples as 0.37 eV and 1.06 eV while DAO, which are the references for GPAO, is 0.37 eV and 1.06 eV respectively.
These result show that GPAO and DAO calculations match well. 
The surface reaction examples demonstrate that GPAO reduces computational cost more than DAO by \emph{three to four orders of magnitude} while giving suppressing the errors of energy barriers under 0.01 eV/atom. 

\section{Multiple conformational transition pathways of alanine dipeptide}

\begin{figure}[htb!]
  \begin{tikzpicture} %
    \node[anchor=south west, inner sep=0] (X) at (0,0){
      \includegraphics[trim=35 50 40 55, width=0.425\textwidth]{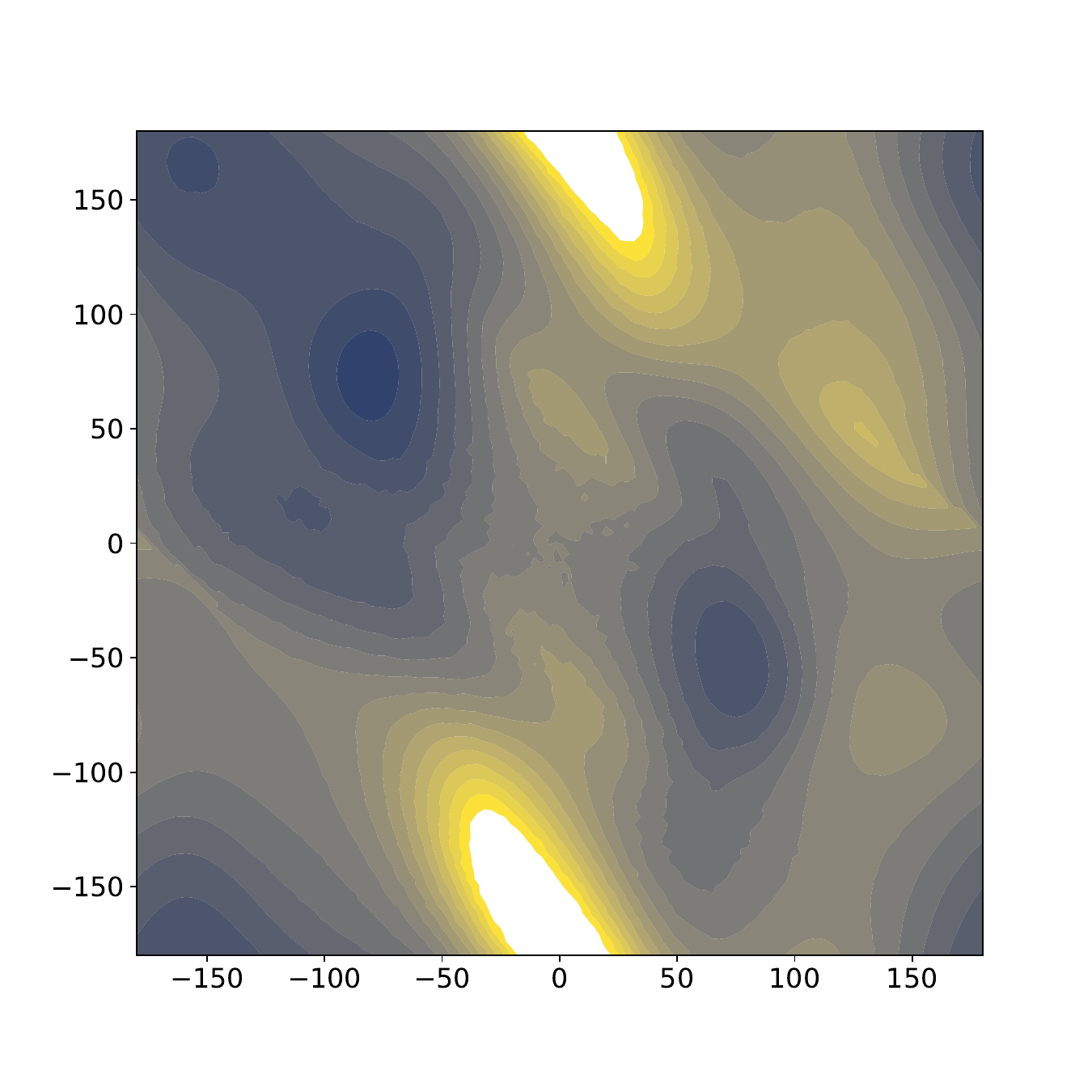}}; %
    \begin{scope}[x={(X.south east)},y={(X.north west)}]
      \node[anchor=north east] (cbar) at (0.5/0.425,1){
        \includegraphics[trim=0 0 27 0, clip, width=0.08\textwidth]{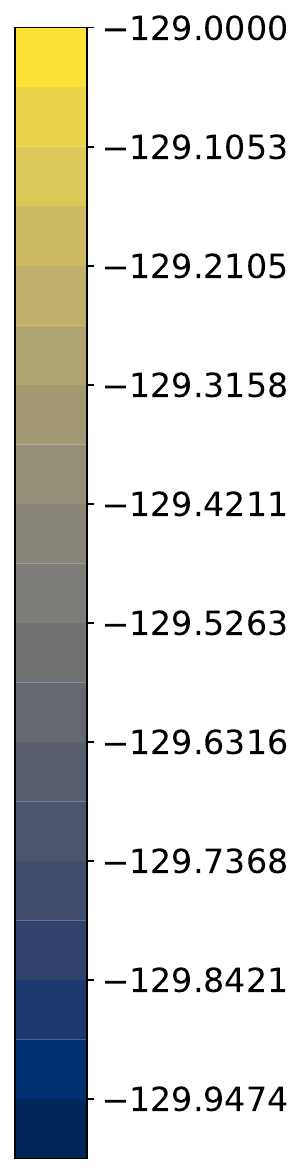}};
      \node[anchor=north west] (C7eq) at (0, 0) {
        \includegraphics[trim=0 0 0 0, width=0.25\textwidth]{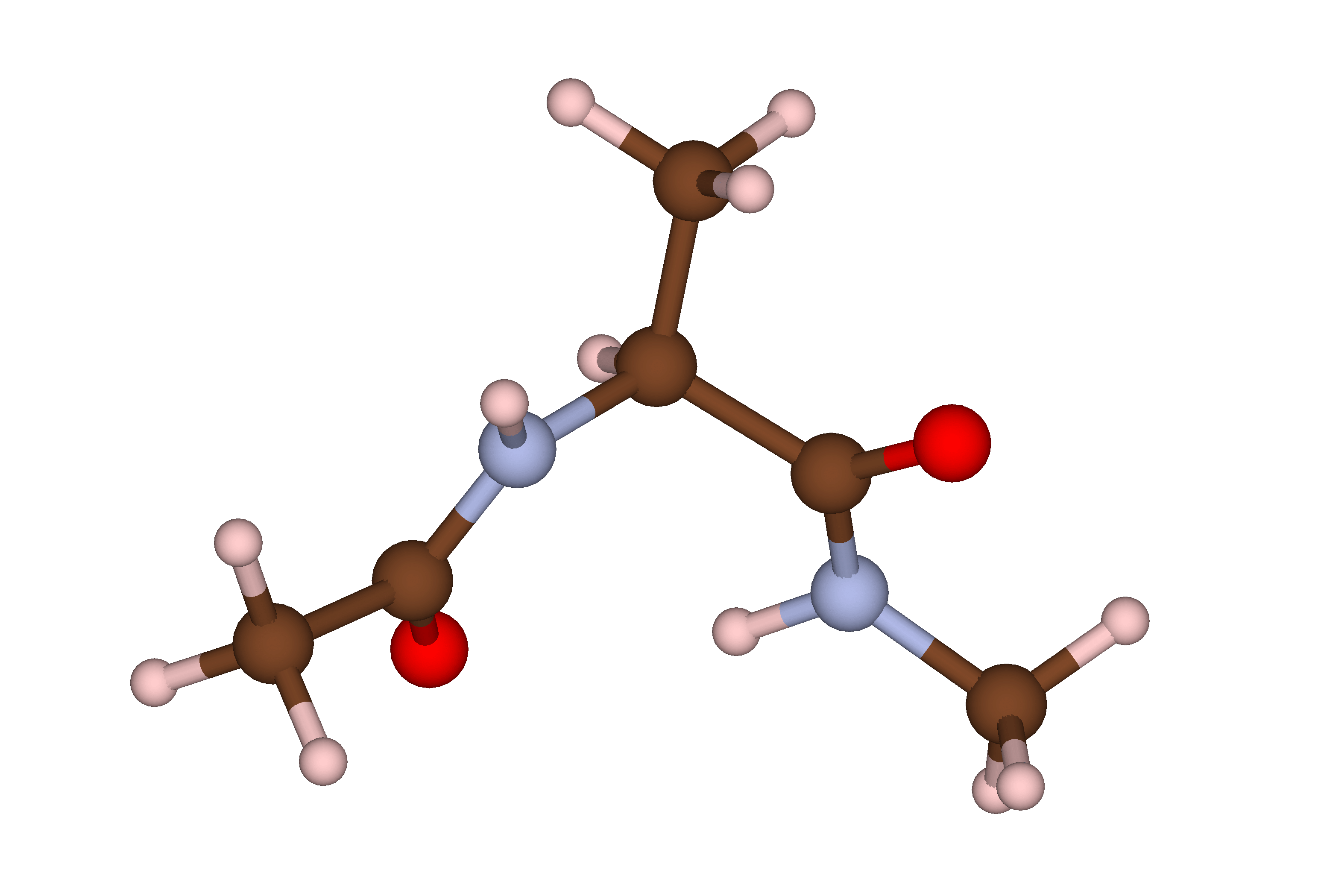}};
      \draw[->, line width=0.25mm] (0.325, 0.678) -> (0.32, 0);
      \node[anchor=north east] (C7ax) at (0.5/0.425, 0) {
        \includegraphics[trim=0 0 0 0 , width=0.25\textwidth]{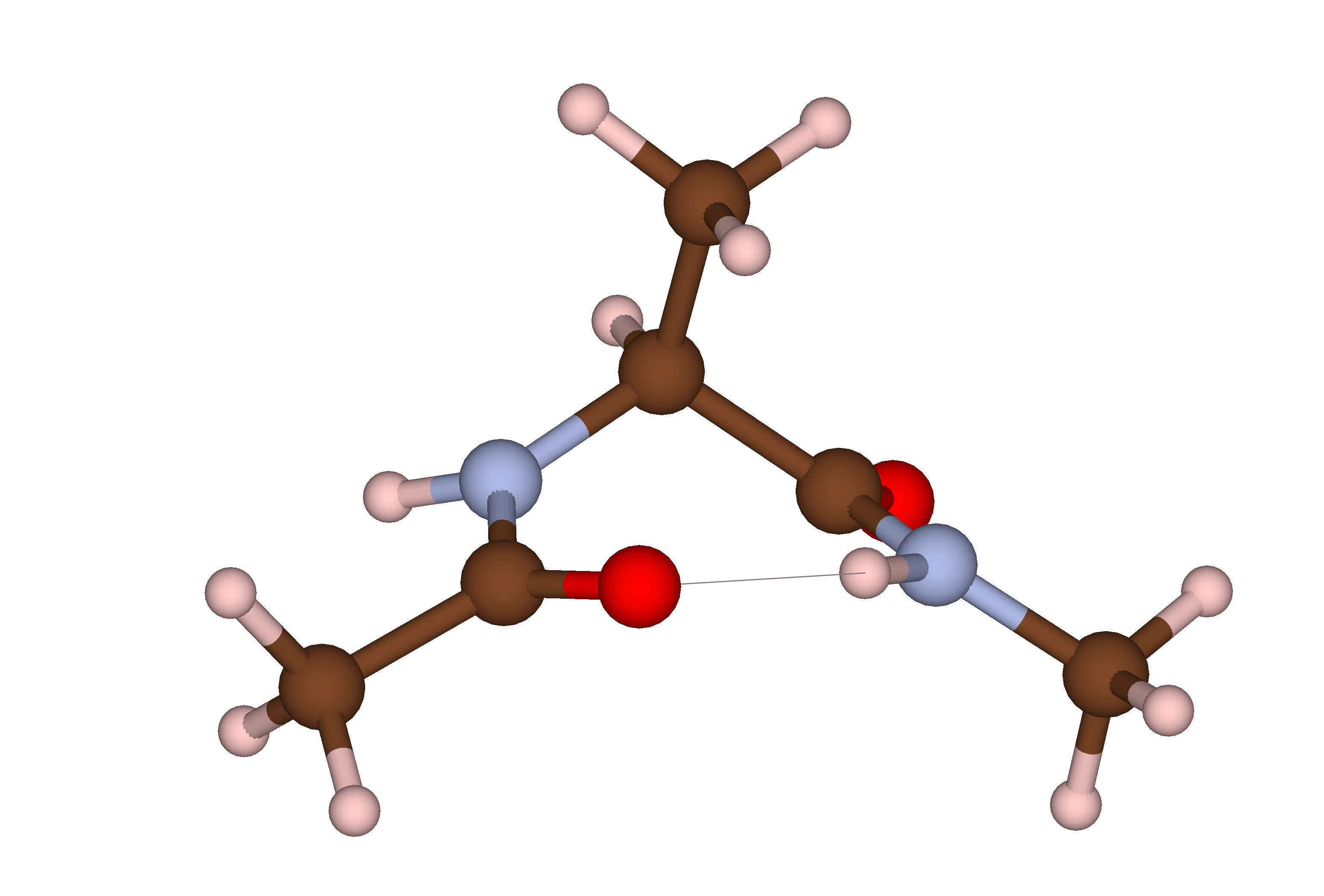}};
      \draw[->, line width=0.25mm] (0.695, 0.38) -> (0.85, 0);
      \node[text width=5in, anchor=south west] (Title) at (0.3,0.98) {Alanine Dipeptide PES (eV)};
      \node[text width=5in, anchor=south west] (a) at (0.,.98) {{(a)}};
      \node[text width=5in, anchor=north west] (b) at (0.,.0) {{(b)}};

      \node[text width=5in, anchor=north west] (C7eq) at (0.0,-0.1) {$C7_{\mathrm{eq}}$};
      \node[text width=5in, anchor=north west] (C7ax) at (0.6,-0.1) {$C7_{\mathrm{ax}}$};
      \node[text width=5in, anchor=north west] (xlabel) at (0.49,0) {$\phi$ (deg)};
      \node[text width=5in, anchor=north west] (ylabel) at (-0.01,0.54) {$\psi$};
      \node[text width=5in, anchor=north west] (phi) at (.179,-.082) {$\phi$};
      \node[text width=5in, anchor=north west] (psi) at (.379,-.082) {$\psi$};

      \node[text width=5in, anchor=north west] (calpha) at (.88,-.15) {$C_{\alpha}$};
      \node[text width=5in, anchor=north west] (cbeta) at (.86,-.02) {$C_{\beta}$};

      \node[text width=5in, anchor=north west] (nl) at (.75,-.15) {$N$};
      \node[text width=5in, anchor=north west] (cl) at (.75,-.3) {$C_{-1}$};

      \node[text width=5in, anchor=north west] (cr) at (0.95,-.18) {$C$};
      \node[text width=5in, anchor=north west] (nr) at (0.93,-.3) {$N_{+1}$};

      \draw (.292,-.25)[red, line width=0.5mm, rotate around={30:(0,0)}, yscale=4] arc(-60:-10:.0125);
      \draw (.2685,-.197)[->,red, line width=0.5mm, rotate around={30:(0,0)}, yscale=4] arc(10:240:.0125);

      \draw (.329,-.255) [red, line width=0.5mm,rotate around={-30:(0,0)}, yscale=4] arc(-60:-10:.0125);
      \draw (.36,-.21) [->, red, line width=0.5mm,rotate around={-30:(0,0)}, yscale=4] arc(10:240:.0125);

    \end{scope}%

  \end{tikzpicture}
  \caption{\label{fig:alanine_map} PES map and structure of alanine dipeptide.
    (a) The PES of alanine dipeptide as a function of reduced coordinates, $\phi$ and $\psi$ angles, $V\left(\psi, \phi\right)$.
    (b) The $C7_{\mathrm{eq}}$ and $C7_{\mathrm{ax}}$ conformations of alanine dipeptide are drawn.
    The brown balls correspond to carbon, the light blue ones are nitrogen, the red ones are oxygen, and the ivory ones are hydrogen.}
\end{figure}
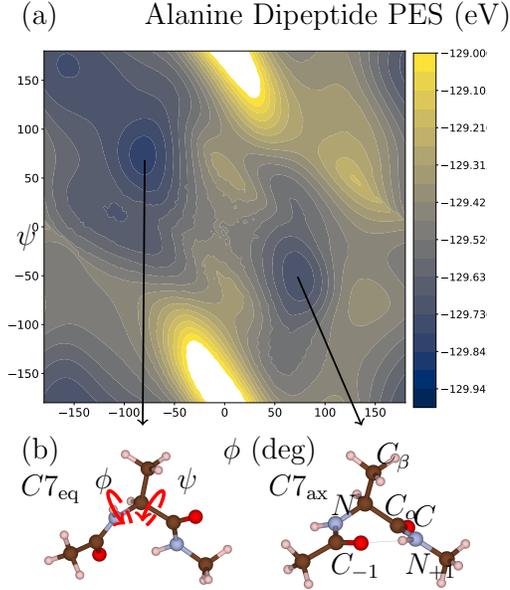

Encouraged by the significant gain of computational efficiency by GPAO in five orders of magnitude,
we investigate multiple conformational transition paths of alanine dipeptide through \textit{ab initio} calculations.
Alanine dipeptide has been widely used as a test system for computational methods in biophysics~\cite{lee_finding_2017}. 
It has two rotatable dihedral angles, the $\phi$ and $\psi$ angles and two stable conformations, $C7_\mathrm{eq}$ and $C7_\mathrm{ax}$ (Figure~\ref{fig:alanine_map}). %, which are accurately represented by the $\phi$ and $\psi$ angles.
The PES's shape shows that multiple conformational transition pathways from one conformation to the other are possible (Figure~\ref{fig:alanine_map}).
%Lee and coworkers discovered eight possible conformational transition pathways through OM action optimization using the Action-CSA method~\cite{lee_finding_2017,lee_direct_2020}.
We combine GPAO with the Action-CSA method that finds multiple transition pathways through the global optimization of action on a trajectory space and call it GP-CSA. 
%A detailed description of the implementation of Action-CSA is given in Supporting Information~\cite{supplement}.
The multiple transition pathways of alanine dipeptide are sampled using GP-CSA calculations with three different total transition times, $t_\mathrm{total} = Ndt =$ 3, 6, and 9 ps. 
Each pathway converged locally using GPAO. 
Further details on the converging process of alanine dipeptide are provided in Figure~\ref{subfig:ADGPAO}. 

\begin{figure}[!ht]
\begin{tikzpicture}%
\node[anchor=south west, inner sep=0] (X) at (0,0){
    \includegraphics[width=0.44\textwidth]{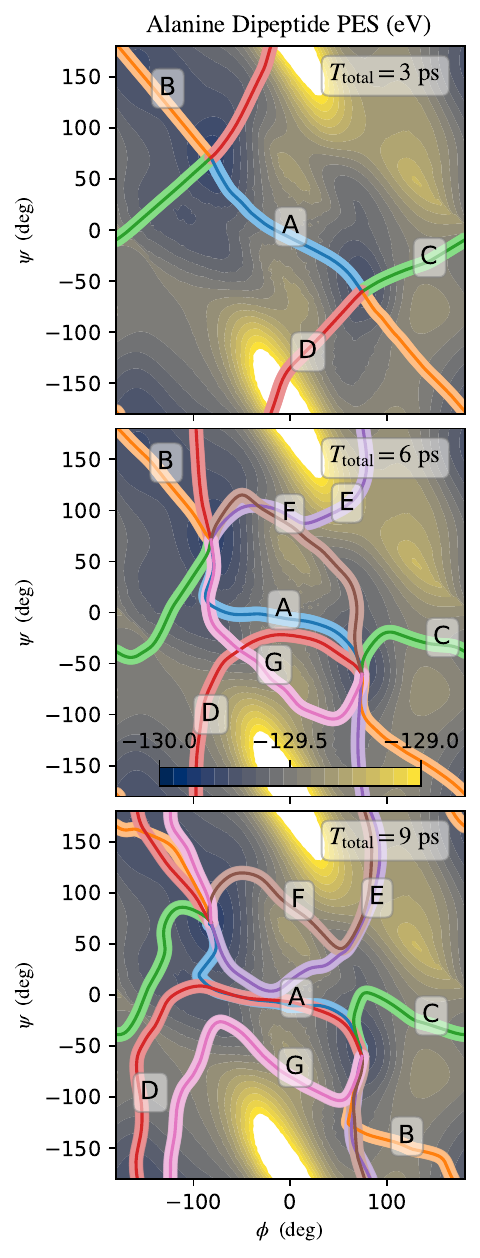}};
\begin{scope}[x={(X.south east)},y={(X.north west)}]%
    \node[text width=5in, anchor=south west, font=\large] (a) at (0.01,.925) {{(a)}};
    \node[text width=5in, anchor=north west, font=\large] (b) at (0.01,.65) {{(b)}};
    \node[text width=5in, anchor=north west, font=\large] (b) at (0.01,.35) {{(c)}};
\end{scope}%
  \end{tikzpicture}
    \caption{The distinctive pathways of each time scale obtained with GP-CSA. 
    Pathways obtained with $t_\mathrm{total} = 30\ \text{\AA}\sqrt{\text{u/eV}}\approx 3\ \text{ps}$ (a), 
    with  $t_\mathrm{total} = 60\ \text{\AA}\sqrt{\text{u/eV}}\approx 6\ \text{ps}$ (b) and 
    with  $t_\mathrm{total} = 90\ \text{\AA}\sqrt{\text{u/eV}}\approx 9\ \text{ps}$ (c). }
    \label{fig:GPCSA}
\end{figure}

The potential energies and forces of alanine dipeptide are calculated using the \textit{ab initio} based density-functional-theory package \textsc{VASP}~\cite{kresse_efficiency_1996, kresse_efficient_1996}. 
The constrained MD method of \textsc{VASP} is used to optimize the conformations of $C7_\mathrm{ax}$ and $C7_\mathrm{eq}$ with fixed $\phi$ and $\psi$ angles.  
Since the structures of alanine dipeptide are almost exclusively defined by $\phi$ and $\psi$ angles, its potential can be expressed as follows:
$V\left(\left\{\mathbf{x}^{\left(n\right)}\right\}\right)=V\left(\phi^{\left(n\right)},\psi^{\left(n\right)}\right)$. 
This reduction on coordinates $ \mathbf{x}^{\left(n\right)}=\left[\phi^{\left(n\right)}, \psi^{\left(n\right)}\right], \mathbf{v}^{\left(n\right)}= \left[\phi^{\left(n+1\right)}-\phi^{\left(n\right)}, \psi^{\left(n+1\right)}-\psi^{\left(n\right)}\right] / {dt}$ is extended on the formulation of the modified OM action for alanine dipeptide. 
Our $\Theta_\mathrm{OM}$ for alanine dipeptide is
\begin{align}
  \Theta_\mathrm{OM} & = S_\mathrm{OM}+\mu_\mathrm{E}\sum_{n=0}^{N-1}\left(E^{\left(n\right)}-E_\mathrm{t}\right)^2 \nonumber \\ 
  & = \frac{\Delta V}{2}+\frac{1}{4}\sum_{n=0}^{N}\left[\frac{dt}{2\gamma}\left(\frac{1}{I_\phi^{\left(n\right)}\left(\phi,\psi\right)}+\frac{1}{I_\psi^{\left(n\right)}\left(\phi,\psi\right)}\right)\left(\left|\nabla V\left(\left\{\mathbf{x}^{\left(n+1\right)}\right\}\right)\right|^2+\left|\nabla V\left(\left\{\mathbf{x}^{\left(n\right)}\right\}\right)\right|^2\right) \right. \nonumber \\
    & -\left(\nabla V\left(\left\{\mathbf{x}^{\left(n+1\right)}\right\}\right)-\nabla V\left(\left\{\mathbf{x}^{\left(n\right)}\right\}\right)\right)\cdot\mathbf{v}^{\left(n\right)} \nonumber \\
    & +\left.\frac{\gamma}{dt}\left(I_\phi^{\left(n\right)}\left(\phi,\psi\right)\left(\frac{\phi^{\left(n+1\right)}-\phi^{\left(n\right)}}{dt}\right)^2+I_\psi^{\left(n\right)}\left(\phi,\psi\right)\left(\frac{\psi^{\left(n+1\right)}-\psi^{\left(n\right)}}{dt}\right)^2\right)\right] \nonumber \\
  & +\mu_\mathrm{E}\sum_{n=0}^{N-1}\left(\frac{1}{2}I_\phi^{\left(n\right)}\left(\phi,\psi\right)\left(\frac{\phi^{\left(n+1\right)}-\phi^{\left(n\right)}}{dt}\right)^2+\frac{1}{2}I_\psi^{\left(n\right)}\left(\phi,\psi\right)\left(\frac{\psi^{\left(n+1\right)}-\psi^{\left(n\right)}}{dt}\right)^2+V\left(\left\{\mathbf{x}^{\left(n\right)}\right\}\right)-E_\mathrm{t}\right)^2,
  \label{eq:s_om2}
\end{align}
%\end{widetext}
where $\nabla V\left(\left\{\mathbf{x}^{\left(n\right)}\right\}\right)=\left[\partial_\phi V\left(\left\{\mathbf{x}^{\left(n\right)}\right\}\right),\ \ \partial_\psi V\left(\left\{\mathbf{x}^{\left(n\right)}\right\}\right)\right]$ indicates the potential difference with respect to $\phi$ and $\psi$, $I_\phi^{\left(n\right)}\left(\phi,\psi\right)$ and $I_\psi^{\left(n\right)}\left(\phi,\psi\right)$ indicate the moments of inertia associated with $\phi$ and $\psi$ angles of $n$'th image, respectively. 
We use the following periodic kernel to describe the PES of alanine dipeptide with GP: 
\begin{equation}
k\left(\mathbf{x}^{ \left( m \right)}, \mathbf{x}^{\left( n \right)} \right) 
    = \sigma_f \exp{\left( -\frac{2}{l^2} \sum_{d=1}^{D}\sin^2 
    \frac{\left( \mathbf{x}_d^{\left(m \right)} - \mathbf{x}_d^{\left(n\right)} \right)}{2} \right),}
\end{equation}
where $\mathbf{x}=(\phi, \psi)$ is a two-dimensional reaction coordinate ($D=2$), $d$ is a dimensional index, and $l, \sigma_f$ are isotropic hyperparameters.

%% Results of alanine dipeptide. 
Among the pathways obtained with GP-CSA, distinctive pathways with low action values are plotted in Figure~\ref{fig:GPCSA}. 
Path {\sffamily A} in Figure~\ref{fig:GPCSA} has the lowest $S_\mathrm{OM}$ value for all transition times, suggesting that Path {\sffamily A} is the most dominant pathway between $C7_\mathrm{eq}$ and $C7_\mathrm{ax}$ conformations. 
This result is consistent with previous studies conducted with molecular mechanics potentials~\cite{lee_finding_2017, elber_temperature_2000}.
The maximum potential energy of a pathway estimated with GP, $\mu^{\mathrm{(max)}}$ is near -128.48 eV (Table. \ref{table:GPCSA}).
For all three different $t_\mathrm{total}$, GP-CSA consistently locates the correct saddle points.

\begin{table}[]
 \caption{Action values and energy barriers of multiple conformational transition pathways of alaninit dipeptide. 
   The OM value and maximum potential energy of conformational transition pathways of alanine dipeptide are obtained with GP-CSA under three different transition time scales.
   The pathways from {\sffamily A} to {\sffamily E} are shown in Figure~\ref{fig:GPCSA}. 
    }
   \label{table:GPCSA}
 %  \begin{ruledtabular}
    \begin{tabular}{cccccccc}
    \hline
    &\multicolumn{3}{c}{$S_\mathrm{OM}$ (eV)} & \multicolumn{3}{c}{$\mu^{\mathrm{(max)}}$ (eV)} \\
    %\hline
    $t_\mathrm{total}$  & 3 ps & 6 ps  & 9 ps & 3 ps & 6 ps  & 9 ps \\
      \hline

      {\sffamily A}& 3.0191 & 3.553 &3.145 & -129.478 &-129.479&-129.482 \\
      {\sffamily B}& 3.8231 & 3.635 &3.505 & -129.387 &-129.449&-129.457 \\
      {\sffamily C}& 3.8913 & 4.124 &3.574 & -129.444 &-129.448&-129.449 \\
      {\sffamily D}& 11.282 & 5.551 &4.887 & -129.203 &-129.416&-129.473 \\
      {\sffamily E}&        & 8.547 &6.745 &          &-129.237&-129.316 \\
      {\sffamily F}&        & 4.561 &6.577 &          &-129.399&-129.315 \\
      {\sffamily G}&        & 5.293 &7.447 &          &-129.366&-129.357 \\

      \hline
    \end{tabular}
\end{table}

The $S_\mathrm{OM}$ value of Path {\sffamily D} with a $t_\mathrm{total}$ of 3 ps is 11.282 eV, which is high enough to make the path improbable.
However, the action value decreases drastically to 5.551 eV and 4.887 eV when $t_\mathrm{total}$ elongates to 6 ps and 9 ps. 
As $t_\mathrm{total}$ increases, Path {\sffamily D} becomes longer, allowing the pathway to detour the high potential region and to seek a lower valley of the PES (see Path {\sffamily D} in Figure~\ref{fig:GPCSA}b and c).  
This tendency is also observed in the $\mu^{\mathrm{(max)}}$ values in Table~\ref{table:GPCSA}. 
$\mu^{\mathrm{(max)}}$ of Path {\sffamily D} at 3 ps is 0.2 eV higher than the other transition times.

It is noticeable that the relatively longer pathways, i.e, Path {\sffamily E}, {\sffamily F}, and {\sffamily G}, are not observed in simulations with $t_\mathrm{total} = 3$ ps.
As $t_\mathrm{total}$ becomes longer, GPAO samples additional pathways, which require longer $t_\mathrm{total}$ to occur.
$S_\mathrm{OM}$ values of Path {\sffamily E} obtained with $t_\mathrm{total}=$ 6 ps and 9 ps are 8.547 eV and 6.745 eV, and $\mu^\mathrm{(max)}$ values are $-129.237$ eV and $-129.316$ eV, respectively. 
The maximum potential energy of Path {\sffamily E} of $t_\mathrm{total} = 6$ ps is 0.1 eV higher than that of Path {\sffamily E} with $t_\mathrm{total} =9$ ps. 
This difference arises because Path {\sffamily E} with $t_\mathrm{total} =6$ ps did not pass the true saddle point due to a short transition time.
However, Path {\sffamily E} with $t_\mathrm{total} = 9$ ps correctly passed the saddle point (see {\sffamily E} in the Figure~\ref{fig:GPCSA}c).

With $t_\mathrm{total}=6$ ps, two additional pathways, Path {\sffamily F} and Path {\sffamily G} are found, which are similar to the most dominant pathway, Path {\sffamily A}.
These two pathways are longer pathways with higher $S_\mathrm{OM}$ values, 4.56 eV and 5.293 eV, than Path {\sffamily A}. 
These results clearly indicate that our GPAO method finds multiple transition pathways of the conformational change of alanine dipeptide accurately with a QM potential, which has not been reported previously.

GPAO also improves the Hessian's accuracy by accumulating data points, which allows one to utilize potential functions whose Hessians are hard to calculate for path sampling. 
The conventional way of obtaining the Hessian of a DFT potential is to utilize a finite difference calculation. 
That is, to acquire a Hessian at an arbitrary point $\phi^{(n)}, \psi^{(n)}$, one needs to compute $\nabla V(\phi^{(n)}, \psi^{(n)}), \nabla V(\phi^{(n)}+\varepsilon, \psi^{(n)})$ and $\nabla V(\phi^{(n)}, \psi^{(n)}+\varepsilon)$. 
These two additional calculations triple the computation cost to evaluate $\Theta_\mathrm{OM}$ preventing the use of finite difference calculation from being applied to relatively large systems.
Thus, a fast evaluation of a potential using GP makes using Hessian for practical problems feasible.

\section{Conclusion}
In this study, we demonstrate that the GP algorithm dramatically enhances the efficiency of searching minimum action pathways by approximating a PES accurately with much-reduced numbers of energy evaluations with diverse examples: the MB potential, isomerization of small molecules, surface reactions, and conformational transition of alanine dipeptide.
For most cases, the transition pathways obtained with GP and DAO show little discrepancies.
The errors of energy barriers are also marginal.
These results indicate that transition pathway search using the modified OM action with a total energy conservation restraint on an approximate PES is an efficient and accurate approach. In addition, compared to the direct optimization of the OM action itself, large fluctuations of kinetic energies are effectively removed without affecting the OM action of a pathway. 

The gradients of OM action require the Hessian calculation of a potential function, which is computationally expensive to obtain in general. 
Thus, combining GP with action optimization gives not only superior computational efficiency also accurate approximations to the gradients of the OM action without actual Hessian calculations.
This advantage is especially significant for the case where the evaluation of potential energies and their Hessians are challenging to obtain, i.e., the Hessian of DFT-based potentials is often incorrect. 
On the contrary, GPAO accumulates gradient information at multiple data points and leads to the accurate estimation of the Hessian from a Gaussian PES. 
Since the Hessian of a Gaussian PES can be estimated accurately, the OM action of a pathway can be readily obtained with DFT-based potentials with enhanced computational efficiency by five orders of magnitude.  
Additionally, we also showed that our GPAO method automatically finds suitable hyperparameters, such as target energy.

This study also demonstrates that combining GP with Action-CSA~\cite{lee_finding_2017} enables the sampling of multiple pathways.
Due to GPAO's efficiency, multiple conformational transition pathways between the two stable conformations of alanine dipeptide are successfully obtained using a DFT-based potential.
This is the first work that sampled multiple low action pathways of alanine dipeptide with a DFT potential to the best of our knowledge. 

Tackling higher-dimensional problems in an efficient way is still a daunting challenge in machine learning fields and other related fields. 
This paper proves that our GPAO method properly works for relatively small-dimensional problems, whose degree of freedom is less than 30. %% PLEASE CHECK!
However, the method does not guarantee its efficiency on problems with a large degree of freedom yet. 
Fortunately, many methods have been suggested to solve problems in a high dimensional space using various types of descriptors~\cite{bartok_representing_2013, behler_atom-centered_2011,behler_generalized_2007,kocer_continuous_2020}.
Furthermore, finding reaction pathways of the systems with many degrees of freedom can be applied to various problems such as finding drug binding pathways, battery, and protein folding. 
Thus, we believe that GPAO will open up new possibilities for such problems.

\section{Methods and Theory}

\subsection{Gaussian Process}
Significant computational improvements of GPAO over the conventional least action methods are attributed to quantifying the variance of specific images in a pathway by using Bayesian inference. 
For example, let us suppose a pathway consisting of $N$ images and $M$ evaluated data points. 
In our GP model, the probability distribution of a given string of images, $\mathbf{X}_\ast=\left[ \mathbf{x}^{\left(1\right)},\mathbf{x}^{\left(2\right)},\cdots, \mathbf{x}^{\left(n\right)},\cdots, \mathbf{x}^{\left(N\right)} \right]$,
whose energies and forces are represented by $\mathbf{Y}_\ast=\left[\boldsymbol{V}_\ast, \nabla\boldsymbol{V}_\ast \right]$, is given by a multivariate Gaussian distribution: 
\begin{equation} p\left(\mathbf{Y}_\ast;\mathbf{X}_\ast,\mathbf{X},\mathbf{Y}\right)
 =\mathcal{N}\left(\mathbf{X}_\ast\middle|\boldsymbol{\mu}_\ast,\ \mathbf{\Sigma}_\ast\right),
\end{equation}
where $\mathbf{X}=\left[ \mathbf{x}^{\left(1\right)},\mathbf{x}^{\left(2\right)},\cdots, \mathbf{x}^{\left(m\right)},\cdots, \mathbf{x}^{\left(M\right)} \right]$ are the evaluated data points and $\mathbf{Y}=\left[\boldsymbol{V}, \nabla{\boldsymbol{V}} \right]$ correspond to their corresponding potential energies and forces. $\boldsymbol{\mu}_\ast$ and $\mathbf{\Sigma}_*$ are the mean and the variance of the Gaussian distribution, respectively:
\begin{align}
\boldsymbol{\mu}_\ast & =
\mathbf{m}_\ast + \mathbf{K}_{\ast}^{T}\left(\mathbf{K}+\sigma_n\mathbf{I}\right)^{-1} \left(\mathbf{Y}-\mathbf{m}\right) \\
\mathbf{\Sigma}_\ast & = \mathbf{K}_{\ast\ast}+\mathbf{K}_\ast^T\left(\mathbf{K}+\sigma_n\mathbf{I}\right)^{-1}\mathbf{K}_\ast,
\end{align}
where $\boldsymbol{m}$ and $\left(\boldsymbol{K}\right)_{mm'}= k\left(\mathbf{x}^{(m)}, \mathbf{x}^{(m')}\right)$ are the mean and the kernel matrix of the multivariate Gaussian distribution, $p(\mathbf{Y})=\mathcal{N}\left(\boldsymbol{m}\left(\mathbf{X}\right),  \boldsymbol{K}\left(\mathbf{X}, \mathbf{X} \right)\right)$. 
$\boldsymbol{m}_\ast$ and $\left(\boldsymbol{K}_{\ast\ast}\right)_{nn'}= k\left(\mathbf{x}^{(n)}, \mathbf{x}^{(n')}\right)$ are the mean and the kernel matrix of $\mathcal{N}\left(\boldsymbol{m}\left(\mathbf{X}_\ast\right), \boldsymbol{K}\left(\mathbf{X_\ast}, \mathbf{X_\ast} \right)\right)$. 
$\left(\mathbf{K}_\ast\right)_{mn}=k\left(\mathbf{x}^{(m)}, \mathbf{x}^{(n)}\right)$ is the kernel matrix connecting coordinates between the evaluated data points and the images in a pathway. 
$\sigma_n$ is a noise parameter and $\mathbf{I}$ is the identity matrix. 
A detailed description of GP is provided in Supporting Information~\cite{supplement}.

\subsection{Classical and Symmetric Onsager-Machlup action with total energy conservation} \label{subsec:Modified classical and Onsager-Machlup}
For a pathway consisting of $N$ images and $A$ atoms for each image, the discretized representation of the classical action with energy conservation restraint, $\Theta_\mathrm{cls}$~\cite{passerone_action-derived_2001}, is
\begin{equation}\label{eq:Scls}
\Theta_\mathrm{cls}\left(\left\{\mathbf{X}_\ast\right\},E_\mathrm{t}\right)=S_\mathrm{cls}+\mu_\mathrm{E}\sum_{n=0}^{N-1}\left(E^{\left(n\right)}-E_\mathrm{t}\right)^2,
\end{equation}
where $\mu_\mathrm{E}$ is the coefficient of the total energy conservation restraint term, $E_\mathrm{t}$ is the target total energy of the system.
The classical action is defined as
\begin{equation}
S_\mathrm{cls} = \sum_{n=0}^{N-1} dt \left[ \frac{1}{2} \sum_{a=0}^{A-1} {\frac{m_{a}  \left(\mathbf{x}_a^{\left(n\right)}-\mathbf{x}_a^{\left(n+1\right)}\right)^2}{dt^2} - V\left(\mathbf{x}^{\left(n\right)}\right)}\right].
\end{equation}
Total energy at the time step $n$ is defined as follows:
\begin{equation}
E^{\left(n\right)}=\frac{1}{2} \sum_{a=0}^{A-1}{\frac {m_{a} \left(\mathbf{x}_a^{\left(n\right)}-\mathbf{x}_a^{\left(n+1\right)}\right)^2}{dt^2} + V\left(\mathbf{x}^{\left(n\right)}\right)}.
\end{equation} 

If a system propagates according to the Langevin dynamics, the dynamics of the diffusive system can be described by the OM action~\cite{machlup_fluctuations_1953a,machlup_fluctuations_1953b,miller2007sampling}.
Miller III and Predescu suggested the symmetric OM action where it is numerically stable up to second order~\cite{miller2007sampling}. For a large system with a highly viscous bathtub, the time evolution of the system is described by 
\begin{equation}
\dot{\mathbf{x}}_t + \gamma^{-1}V'(\mathbf{x}_t) = \dot{\mathbf{W}}_t,
\end{equation}
where $\dot{\mathbf{x}}$ is time derivative of the coordinates, $\gamma$ is friction coefficient, and $\dot{\mathbf{W}}$ is time derivative of the Wiener process or Brownian motion. Transition probability of the distribution of system is
\begin{equation}
    P_{AB}(\mathbf{x}_0, \mathbf{x}_t; \Delta t) = I_A(\mathbf{x}_0) P(\mathbf{x}_0,\mathbf{x}_t; \Delta t) I_B(\mathbf{x}_t),
\end{equation}
where $I$ is an indicator function of configuration space. From the Markovian properties of diffusive dynamics, the Canonical ensemble in the form of re-weighted propagator is,
\begin{equation}
    P(\mathbf{x}_0, \mathbf{x}_t, \Delta t) = e^{-\beta\frac{V(\mathbf{x}_t)- V(\mathbf{x}_0)}{2}} \prod_{k=0}^{n-1} G(\mathbf{x}_k, \mathbf{x}_{k+1}, t),
\end{equation}
where $t = \Delta t / n$. $G(\mathbf{x}_k, \mathbf{x}_{k+1}, t)$ follows Bloch equation,
\begin{equation}\label{eq:felixbloch}
   \frac{\partial}{\partial t} G(\mathbf{x}', \mathbf{x}, t)  = D\left[\frac{\partial^2}{\partial \mathbf{x}'^2} - V_{\mathbf{eff}}\right] G(\mathbf{x}', \mathbf{x}, t),
\end{equation}
where $V_{\mathbf{eff}}$ is effective potential written as, 
\begin{equation}
    V_{\mathbf{eff}} = \frac{D\beta^2}{4} V'{^2}(\mathbf{x}) - \frac{\beta D}{2} V''(\mathbf{x}).
\end{equation}
Solution for the Eq.~\ref{eq:felixbloch} is Feynman-Kac equation,
\begin{equation}
    G(x',x;t) = p_{\sigma^2}(x',x) \mathbb{E}^{\sigma^2}_{x',x} \exp{\left(\frac{-\beta^2 D}{4} \int_0^t V'(W_{t'})dt' + \frac{\beta D}{2} \int_0^t V''(W_{t'})dt'\right)},
\end{equation}
where $p_{\sigma^2}(x',x) = (2\pi \alpha^2)^{-1/2} e^{-(x'-x)^2/(2\alpha^2)}$ and $\mathbb{E}^{(\sigma^2)}_{x',x} $ is symbol for average of all Brownian motion with variance $\sigma^2=2Dt$. To alleviate the second order derivation term Miller III and Predescu suggested dividing $ \int_0^t V''(W_{t'})dt'$ into two finite difference of first order terms~\cite{miller2007sampling}. That is,
\begin{align}
& 2D\int_0^t V''(W_{t'})dt' \notag\\
&\approx \lim_{\Delta t \rightarrow 0} \sum_{k=1}^n V''(W_{t_k})2D\Delta t_k \notag \\
&= \lim_{\Delta t \rightarrow 0} \sum_{k=1}^n V''(W_{t})[W_{t_{k+1}} - W_{t_k}]^2 \notag \\
&= \lim_{\Delta t \rightarrow 0} \sum_{k=1}^n [V'(W_{t_{k+1}}) - V'(W_{t_k})][W_{t_{k+1}} - W_{t_k}].
\end{align}

We get discrete propagator for diffusive dynamics without containing Hessian,
\begin{align}
    & G(x', x; t) \approx p_{\sigma^2}(x', x) \mathbb{E}^{(\sigma^2)}_{x',x} \notag\\ 
    & \exp{\left( - \frac{\beta^2D}{4} \sum_{k=0}^n w_kV'^{2}(W_k) + \frac{\beta}{4} \sum_{k=0}^n ([V'(W_k) - V'(W_{k-1})][W_k-W_{k-1}]) \right)}.
\end{align}

Transition probability that states $A$ at time $0$ changes into state $B$ at time $t$ is,
 \begin{equation}
    P_{AB}(\mathbf{x}_0, \mathbf{x}_n; t) = I_A(\mathbf{x}_0)e^{-(\beta/2)V(\mathbf{x}_0)}\left[\prod_{k=1}^n {G_0}\left(\mathbf{x}_k, \mathbf{x}_{k-1}; t/n\right)\right]I_B(\mathbf{x}_n)^{-(\beta/2)V(\mathbf{x}_n)}. 
 \end{equation}

where
 \begin{equation}
    G_0(x', x;\Delta t) = p_{\sigma^2/n} \exp\left(\left\{ - \frac{\beta}{8\gamma} (V'^{2}(x')+ V'^{2}(x)) + \frac{\beta}{4} ([V'(x) - V'(x')](x-x')). \right\}\right) 
 \end{equation}
More detailed derivation can be found in Ref~\cite{miller2007sampling}. The discretized formula of the OM action is given by 
\begin{align}\label{eq:Som}
S_\mathrm{OM} = \frac{\Delta V}{2} + & \frac{1}{4} \sum_{n=0}^{N}\left[\frac{dt}{2\gamma}\left(\left|\nabla V\left( \mathbf{x}^{\left(n+1\right)} \right)\right|^2+\left|\nabla V\left( \mathbf{x}^{\left(n\right)} \right)\right|^2\right) \right. \notag \\
 & \left.  - \left(\nabla V\left( \mathbf{x}^{\left(n+1\right)} \right) - \nabla V \left( \mathbf{x}^{\left(n\right)} \right)\right) \cdot \mathbf{v}^{\left(n\right)}+\frac{\gamma}{dt}\mathbf{v}^{\left(n\right)\mathbf{T}} \cdot \mathbf{v}^{\left(n\right)} \right]. 
\end{align}

Here, $\Delta V$ is the energy difference between the initial and the final step, and $\gamma$ is the damping constant. 
In this work, we use the modified OM action with the total energy conservation restraint:
\begin{equation}
\Theta_\mathrm{OM}\left(\left\{\mathbf{X}_\ast\right\}, E_\mathrm{t} \right) = S_\mathrm{OM}+\mu_\mathrm{E}\sum_{n=0}^{N-1}\left(E^{\left(n\right)}-E_\mathrm{t}\right)^2.
\end{equation}

\subsection{Gaussian Process Action Optimize: GPAO}
In this work, we also implement a routine that automatically finds suitable target energy of a pathway during pathway sampling. 
Target energy is a parameter that is inherently hard to know prior to the sampling of an actual pathway because the proper target energy is closely related to the height of the energy barrier of a pathway.

Our GPAO method handles two tasks simultaneously. 
First, it minimizes the uncertainty of a Gaussian PES near the sampled pathways in an iterative way. 
Second, the method determines the total energy ($E_\mathrm{t}$) of the MEP automatically. 

In this study, we assume that the $E_\mathrm{t}$ of a system is close to the maximum potential energy along with the MEP, $V^{\mathrm{(max)}}$.
To gradually approximate the $E_\mathrm{t}$ to $V^{\mathrm{(max)}}$, we set $E_\mathrm{t}$ as $\left(\boldsymbol{\mu}^{\mathrm{(max)}}_{\ast}-E_\mathrm{t}-0.05\right)/2 $, where $\boldsymbol{\mu}^{\mathrm{(max)}}_{\ast}$ is the estimated maximum potential energy of MEP on the Gaussian PES. 

\subsubsection{Computational details for the M\"uller-Brown potential}
% ### parameters MB ###
Both GPAO and DAO use the same number of intermediate images of 300.
To ensure the continuity of a pathway, we use 300 images for a pathway for both methods. %, with the previous study~\cite{passerone_action-derived_2001} on the MB potential. 
Thus, the energy and forces of 300 images must be calculated for a single action evaluation.
Total transition time is set to 3 ps, damping constant, $\gamma$, is 1, and energy conservation restraint constant, $\mu_{E} = 1$. 
For GPAO, we used the squared exponential kernel for the Gaussian kernel:
\begin{equation}
  \left(\mathbf{x}^{\left(i\right)},\mathbf{x}^{\left(j\right)}\right) = \sigma_f\exp{\left(-\frac{1}{2l^2}\left(\mathbf{x}^{\left(i\right)}-\mathbf{x}^{\left(j\right)}\right)^T\left(\mathbf{x}^{\left(i\right)}-\mathbf{x}^{\left(j\right)}\right)\right)},
\end{equation}
where $l$ and $\sigma_f$ are the isotropic real hyperparameters. 
We use the multivariate Gaussian distribution with zero mean for GPAO calculations, $\mathbf{m}=\mathbf{m}_\ast=0$. 
For DAO, target energy is set to $-0.368$, estimated from the GPAO calculations.
%This value is the results we get from GPAO with the iterative target energy converging process.

\subsubsection{Parameters for isomerization of organic molecules}
% ### Parameters Mol
The DFT calculations are performed with the \textsc{VASP} package~\cite{kresse_efficiency_1996, kresse_efficient_1996}.
The PBE exchange functional, energy cutoff of $520$ eV, and two-body van der Waals interaction are used. 
For DAO and GPAO calculations, OM action with energy restraint action $\left(\Theta^\mathrm{OM}\right)$ is used, $\gamma = 1.0$, $\mu_E = 1.0$. 
Total transition time is set to 0.1 ps and the 150 intermediate images are used, $N=150$, but only 100 sine components, $N_k=100$, are used for action optimization.
For GPAO, we use the standard kernel with isotropic hyperparameters, $\sigma_f$, $l$. 
The ranges of hyperparameters are set to $0.001 < \sigma_f  < 1000$, $0.01 < l^2 < 10^{12}$, $0.0001 < \sigma_n^e < 0.01$, and $0.00001 < \sigma_n^f < 0.001$. 
In this case, using zero-mean, as we used in the MB potential, will poorly fit the PES since the target energy is aggregated near $-20~$ eV to $-35$ eV. For this reason, we use `maximum constant mean', where the maximum potential energy of the MEP of the previous iteration, $\mathbf{\mu}^{(\mathrm{max})}$ is assumed as a mean of Gaussian distribution.  %% WHAT DOES THIS MEAN?
For DAO, target energy has used the value attained by the GPAO counterpart and the results of the GPAO calculations are used as the initial trajectories of the DAO calculations.

\subsubsection{Parameters for surface reactions}
% ###### Parameters Surface
The calculator we used for this example is the embedded medium theory (EMT) calculator implemented under the Atomic Simulation Environment package (ASE).
Parameters of OM action with energy restraint, $\gamma = 1$, $\mu_E = 1$, and $E_t$ are set by the value obtain by GPAO. 
Both GPAO and DAO methods conducted straight line conditions, total transition time is set to $10$ ps, a number of the intermediate images, $N$, is $150$, and a number of sine components used in action optimization are 50, $N_k = 50$.
For DAO, Hessian is calculated using the central finite difference method. 
For GPAO, the standard kernel with the `max mean' function, where `max mean' is the mean function that returns the maximum potential energy of the previous iteration.  
The range of hyperparameters are set as $0.001 < \sigma_f < 1000$, $0.01 < l^2 < 10$, $10^{-6} < \sigma_n^e < 0.01$, and $10^{-7} < \sigma_n^f < 0.001$.

\section{Code and Data availability}

The source code of GPAO can be found in the GitHub repository \\
(https://github.com/schinavro/taps).
GPAO consists of multiple tools in the open-source package \textsc{TAPS}. 
\textsc{TAPS} incorporates tools needed for data-driven pathway search methods such as a database or atomic calculator. 
For atomic calculations, we use the atomic simulation environment (\textsc{ASE}) ~\cite{larsen_atomic_2017} package, which can easily link with atomic calculators such as \textsc{VASP}~\cite{kresse_efficiency_1996, kresse_efficient_1996}, \textsc{Quantumespresso}~\cite{quantum_espresso_giannozzi_quantum_2009}, \textsc{OpenMX}~\cite{openmx_ozaki_efficient_2005}, and SIESTA~\cite{siesta_soler_siesta_2002}.

\begin{acknowledgement}
We like to thank to \censor{In-Ho Lee} for the help in the initial stage of this project. 
\censor{JL} acknowledges the support of the \censor{National} \censor{Research} \censor{Foundation} \censor{of} \censor{Korea} \censor{(NRF)} grant funded by \censor{the Korean government} \censor{(MSIT)} \censor{(No. 2018R1C1B600543513)}.
\censor{JY} acknowledges the support by the \censor{National} \censor{Research} \censor{Foundation} \censor{of} \censor{Korea} \censor{(NRF)}(No. \censor{2020R1F1A1066548}). Additional financial support from \censor{Samsung} \censor{Electronics} is also acknowledged.
This work was also supported by the \censor{National} \censor{Supercomputing} \censor{Center} with \censor{supercomputing} \censor{resources} and \censor{technical supports} \censor{(KSC-2020-CRE-0026)}.
\end{acknowledgement}

\bibliography{main}

\begin{thebibliography}{10}
\bibitem{supplement}
{See} {Supporting} {Information} at [url will be inserted by publisher] for
  detailed explanation on {Gaussian} process regression.
\end{thebibliography}

\end{document}

\section{Discretized formula for the derivatives of modified action at the time step $n$}\label{sisec:discretized}
Discretized formula for the classical action and energy restraint terms are 
\begin{equation}
\frac{\partial S_\mathrm{cls}}{\partial\mathbf{x}^{\left(n\right)}} = dt\sum_{a=0}^{A-1}\left(-\frac{\partial}{\partial\mathbf{x}_a^{\left(n\right)}}V\left(\mathbf{x}^{\left(n\right)}\right)+m_a\mathbf{a}_a^{\left(n\right)}\right) 
\end{equation}

\begin{align}
\frac{\partial}{\partial\mathbf{x}^{\left(n\right)}}& \left(\mu_\mathrm{E}\sum_{n=0}^{N-1}\left(E^{\left(n\right)}-E_\mathrm{t}\right)^2\right) \notag
\\
& = 2\mu_\mathrm{E} dt\sum_{a=0}^{A-1}{\left[ \left(\frac{\partial}{\partial\mathbf{x}_a^{\left(m\right)}}V\left( \mathbf{x}^{\left(n\right)} \right) + \frac{1}{dt}\mathbf{p}_a^{\left(m\right)}\right)\right.
 \left(-E_\mathrm{t} + V\left(\mathbf{x}^{\left(n\right)}\right) + \frac{1}{2} m_a \mathbf{v}_a^{\left(n\right)T} \cdot \mathbf{v}_a^{\left(n\right)}\right)} 
\\
& \left.+ \frac{1}{dt} \mathbf{p}_a^{\left(n-1\right)}\left(-E_\mathrm{t}+V\left(\mathbf{x}^{\left(n-1\right)} \right) + \frac{1}{2} m_a \mathbf{v}_a^{\left(n-1\right)T} \cdot \mathbf{v}_a^{\left(n-1\right)}\right)\right],\notag
\end{align}
where $\mathbf{v}_a^{\left(n\right)}=(\mathbf{x}_a^{\left(n+1\right)}-\mathbf{x}_a^{\left(n\right)})/dt, \mathbf{a}_a^{\left(n\right)}=(2\mathbf{x}_a^{\left(n\right)}-\mathbf{x}_a^{\left(n-1\right)}-\mathbf{x}_a^{\left(n+1\right)})/dt^2$ and $\mathbf{p}_a^{\left(n\right)}=m_a\mathbf{v}_a^{\left(n\right)}$. 
Our parameter set for the classical action on the MB potential is $dt = 0.02$, $m_a=1$, $\mu_\mathrm{E} = 0.1$, $E_\mathrm{t}=-0.45$ and $N=300$.

 Discretized formula for the derivatives of OM action is
\begin{align}
    \frac{\partial S_\mathrm{OM}}{\partial\mathbf{x}^{\left(n\right)}} &= \frac{\gamma dt}{2}\mathbf{a}^{\left(n\right)}\notag \\ 
& +\frac{\partial\nabla V\left( \mathbf{x}^{\left(n\right)} \right)}{\partial\mathbf{x}^{\left(n\right)}}\left(\frac{dt}{2\gamma}\nabla V \left( \mathbf{x}^{\left(n\right)} \right) 
- \frac{dt^2}{4}\mathbf{a}^{\left(n\right)}\right)\notag \\ 
& +\frac{1}{4}\nabla V\left(\mathbf{x}^{\left(n-1\right)} \right)-\frac{1}{2}\nabla V\left(\mathbf{x}^{\left(n\right)} \right) + \frac{1}{4}\nabla V\left(\mathbf{x}^{\left(n+1\right)}\right),
\end{align}
where $\partial \nabla V\left(\mathbf{x}^{\left(n\right)} \right) / \partial\mathbf{x}^{\left(n\right)}$ is the Hessian. 

\section{Details on Gaussian Process Regression}
\label{supsec:Gaussian Process regression}

For an atomic system, the vector representation for the input coordinates of data is

\begin{align} \label{seq:d=3a}
\mathbf{x}^{\left(m\right)} =& \left(x_1^{\left(m\right)},x_2^{\left(m\right)},\cdots,x_a^{\left(m\right)}, \cdots,x_A^{\left(m\right)}, \right.\notag \\
 & y_1^{\left(m\right)},y_2^{\left(m\right)},\cdots,y_a^{\left(m\right)}, \cdots,  y_A^{\left(m\right)},\notag \\
 & \left. z_1^{\left(m\right)},z_2^{\left(m\right)},\cdots,z_a^{\left(m\right)}, \cdots, z_A^{\left(m\right)}\right),
\end{align}
where $x_a^{\left(m\right)}, y_a^{\left(m\right)}, z_a^{\left(m\right)}$ are the $x, y, z$ coordinates of the $a$-th atom of the $m$-th image. 
For an arbitrary coordinate system with $D$ dimensions, the representation vector for the system is simply
\begin{align}
  \mathbf{x}^{\left(m\right)}=\left(x_1^{\left(m\right)},\ x_2^{\left(m\right)},\cdots,x_d^{\left(m\right)}, \cdots,\ x_D^{\left(m\right)}\right),
\end{align}
where $x_d^{\left(m\right)}$ is the element of $d$-th coordinate at $m$-th image. For example, an atomic system (Eq.~(\ref{seq:d=3a})) has $D=3A$ and M\"uller-Brown potential has $D=2$.

The representation for the entire dataset of GP or pathways can be expressed in a matrix form by combining all the data into one where each datum vector is concatenated laterally. 
With the $M$ number of evaluated data points and the $N$ number of images, representation for data points and pathway $\mathbf{X}$ and $\mathbf{X}_\ast$ have the size $\left(D\times M\right)$, $\left(D\times N\right)$, respectively:

\begin{align}
 \mathbf{X} =& \sum_{m=1}^{M}{\mathbf{e}^{\left(m\right)}\otimes\mathbf{x}^{\left(m\right)}} =  \left( \mathbf{x}^{\left(1\right)}\right.\  \mathbf{x}^{\left(2\right)} \cdots  \left.\mathbf{x}^{\left(M\right)} \right) \\
 \mathbf{X}_\ast =& \sum_{n=1}^{N}{\mathbf{e}^{\left(n\right)}\otimes\mathbf{x}^{\left(n\right)}} =  \left(\mathbf{x}^{\left(1\right)}\right.\  \mathbf{x}^{\left(2\right)} \cdots  \left.\mathbf{x}^{\left(N\right)}\right).
\end{align}

The representation for the output $\mathbf{Y}$ has the $\left(D+1\right)M$ dimensional vector with the following sequence:
 
\begin{align}
 \mathbf{Y}= & \left(V^{\left(1\right)},V^{\left(2\right)},\cdots,V^{\left(M\right)}, \frac{\partial V^{\left(1\right)}}{\partial x_1^{\left(1\right)}},\frac{\partial V^{\left(2\right)}}{\partial x_1^{\left(2\right)}},\cdots,\frac{\partial V^{\left(M\right)}}{\partial x_1^{\left(M\right)}},\right. \notag\\
& \frac{\partial V^{\left(1\right)}}{\partial x_2^{\left(1\right)}},\frac{\partial V^{\left(2\right)}}{\partial x_2^{\left(2\right)}},\cdots,\frac{\partial V^{\left(M\right)}}{\partial x_2^{\left(M\right)}}, \left. \frac{\partial V^{\left(1\right)}}{\partial x_3^{\left(1\right)}},\cdots,\frac{\partial V^{\left(M\right)}}{\partial x_D^{\left(M\right)}}\right),
 \end{align}
where $V^{\left(m\right)}$ is the potential energy of the $m$-th image  $\mathbf{x}^{\left(m\right)}$ and $\partial V^{\left(m\right)}/\partial x_d^{\left(m\right)}$ is the gradient of the potential with respect to $x_d^{\left(m\right)}$. 
 
The covariance or kernel function, $k\left(\mathbf{x}^{\left(m\right)},\mathbf{x}^{\left(n\right)}\right)$,  quantifies how two data $\mathbf{x}^{\left(m\right)}$ and $\mathbf{x}^{\left(n\right)}$ are related. 
In the case of the MB potential example, we use the standard (squared exponential) kernel. For the alanine dipeptide example, which has a periodic potential energy surface (PES), we use a periodic kernel. 
Our method also incorporates gradient information to exploit both potential energies and the gradient of system energies to construct a more accurate Gaussian PES. 
The elements of a simple kernel matrix $\boldsymbol{K}$ can be written as 
\begin{align}
   {(\boldsymbol{K})}_{mn}=k\left(\mathbf{x}^{\left(m\right)},\mathbf{x}^{\left(n\right)}\right),
\end{align}
where $(m,\ n)\ \in \left( M\times N\right)$ are the indices of the matrix. 
$\partial_{\mathit{n}} \boldsymbol{K}$,$\ \partial_{\mathit{m}}\boldsymbol{K}$ and $\partial_{\mathit{mn}}\boldsymbol{K}$ are the tensors derived from the derivatives of the simple kernel matrix with their shapes: $\left(M\times D\times N\right),\ \left(D\times M\times N\right)$, and $\left(D\times M\times D\times N\right)$, respectively. 
The components for these tensors are expressed as follows: 
\begin{align}
    \left(\partial_n\boldsymbol{K}\right)_{m,d,n}=\frac{\partial}{\partial x_d^{\left(n\right)}}k\left(\mathbf{x}^{\left(m\right)},\mathbf{x}^{\left(n\right)}\right)\\\left(\partial_m\boldsymbol{K}\right)_{d,m,n}=\frac{\partial}{\partial x_d^{\left(m\right)}}k\left(\mathbf{x}^{\left(m\right)},\mathbf{x}^{\left(n\right)}\right)\\\left(\partial_{mn}\boldsymbol{K}\right)_{d,m,d^\prime,n}=\frac{\partial}{\partial x_d^{\left(m\right)}}\left(\frac{\partial}{\partial x_{d^\prime}^{\left(n\right)}}k\left(\mathbf{x}^{\left(m\right)},\mathbf{x}^{\left(n\right)}\right)\right).
\end{align}

An extended kernel matrix $\boldsymbol{K}_{\mathrm{ext}}$ can be constructed via reshaping its tensors into a matrix form. We reshape the tensors $\partial_n\boldsymbol{K},\ \partial_m\boldsymbol{K}$ and $\partial_{mn}\boldsymbol{K}$ of the form $\left(M\times D\times N\right)$, $\left(D\times M\times N\right)$, $\left(D\times M\times D\times N\right)$ into $\left(M\times DN\right),\ \left(DM\times N\right),\ \left(DM\times DN\right)$. The explicit expression can be written as follows:
%\begin{widetext}
\begin{equation*}
    \boldsymbol{K}_{\mathrm{ext}} = \left(
 \begin{matrix} \boldsymbol{K}&\partial_n\boldsymbol{K}\\ 
 \partial_m\boldsymbol{K}&\partial_{mn}\boldsymbol{K} 
\end{matrix}\right) = 
\left(\begin{matrix}
\kernel{\mathbf{X}}{\mathbf{X'}}& \dkernel{\mathbf{X}}{\mathbf{x'}^\mathrm{(1)}}{\mathbf{x'}^{(1)}}&\dkernel{\mathbf{X}}{\mathbf{x'}^\mathrm{(2)}}{\mathbf{x'}^{(2)}}&\cdots&\dkernel{\mathbf{X}}{\mathbf{x'}^\mathrm{(N)}}{\mathbf{x'}^\mathrm{(N)}}
\\
\dkernel{\mathbf{x}^\mathrm{(1)}}{\mathbf{X'}}{\mathbf{x}^{(1)}}&\ddkernel{\mathbf{x}^\mathrm{(1)}}{\mathbf{x'}^\mathrm{(1)}}{\mathbf{x}^{(1)}}{\mathbf{x'}^{(1)}}&\ddkernel{\mathbf{x}^\mathrm{(1)}}{\mathbf{x'}^\mathrm{(2)}}{\mathbf{x}^{(1)}}{\mathbf{x'}^{(2)}}&\cdots&\ddkernel{\mathbf{x}^\mathrm{(1)}}{\mathbf{x'}^\mathrm{(N)}}{\mathbf{x}^{(1)}}{\mathbf{x'}^\mathrm{(N)}}\\
\dkernel{\mathbf{x}^\mathrm{(2)}}{\mathbf{X'}}{\mathbf{x}^{(2)}}&\ddkernel{\mathbf{x}^\mathrm{(2)}}{\mathbf{x'}^\mathrm{(1)}}{\mathbf{x}^{(2)}}{\mathbf{x'}^{(1)}}&\ddkernel{\mathbf{x}^\mathrm{(2)}}{\mathbf{x'}^\mathrm{(2)}}{\mathbf{x}^{(2)}}{\mathbf{x'}^{(2)}}&\cdots&\ddkernel{\mathbf{x}^\mathrm{(2)}}{\mathbf{x'}^\mathrm{(N)}}{\mathbf{x}^{(2)}}{\mathbf{x'}^\mathrm{(N)}}\\
\vdots&\vdots&\vdots&\ddots&\\
\dkernel{\mathbf{x}^\mathrm{(M)}}{\mathbf{X'}}{\mathbf{x}^\mathrm{(M)}}&\ddkernel{\mathbf{x}^\mathrm{(M)}}{\mathbf{x'}^\mathrm{(1)}}{\mathbf{x}^\mathrm{(M)}}{\mathbf{x'}^{(1)}}&\ddkernel{\mathbf{x}^\mathrm{(M)}}{\mathbf{x'}^\mathrm{(2)}}{\mathbf{x}^\mathrm{(M)}}{\mathbf{x'}^{(2)}}& &\ddkernel{\mathbf{x}^\mathrm{(M)}}{\mathbf{x'}^\mathrm{(N)}}{\mathbf{x}^\mathrm{(M)}}{\mathbf{x'}^\mathrm{(N)}}
\end{matrix}\right),
\end{equation*}
%\end{widetext}
where the shape of the matrix  $\kernel{\mathbf{X}}{\mathbf{X'}}= \boldsymbol{K}$ is $\left(M\times N\right)$, $ \partial \kernel{\mathbf{X}}{\mathbf{x'}^\mathrm{(n)}} / \partial \mathbf{x'}^{(n)}$ is $\left(M\times D\right)$, $  \partial \kernel{\mathbf{x}^\mathrm{(m)}}{\mathbf{X'}} / \partial \mathbf{x}^{(m)}$ is $\left(D\times N\right)$, and $ \partial^2\kernel{\mathbf{x}^\mathrm{(m)}}{\mathbf{x'}^\mathrm{(n)}} / \partial {\mathbf{x}^\mathrm{(m)}}\partial{\mathbf{x'}^\mathrm{(n)}} $is $\left(D \times D\right)$ matrix. In total, the shape of the matrix $\boldsymbol{K}_\mathrm{ext}$ is $\left(D(M+1)\times D(N+1)\right)$.

Bayesian inference comes handy if one wants to estimate the energies of unknown pathway $\mathbf{X}_\ast$ when the observed energy $\mathbf{Y}$ of a structure $\mathbf{X}$ is given. Specifically, the probability of $\mathbf{X}_\ast$ to have $\mathbf{Y}_\ast$, for given $\mathbf{X}$ and $\mathbf{Y}$ can be calculated by integrating throughout possible posterior.
Since the integration of Gaussian is also Gaussian, we obtain a Gaussian function with posterior predictive mean $\boldsymbol{\mu}_\ast$ and covariance $\mathbf{\Sigma}_\ast$ as follows:
\begin{align}
 p\left(\mathbf{Y}_\ast | \mathbf{X}_\ast,\mathbf{X},\mathbf{Y}\right) = &\int{p\left(\mathbf{Y}_\ast|\mathbf{X}_\ast,\mathbf{w}\right)p\left(\mathbf{w}|\mathbf{X},\mathbf{Y}\right)d\mathbf{w}}\notag \\
 = &\mathcal{N}\left(\mathbf{X}_\ast\middle|\boldsymbol{\mu}_\ast,\ \mathbf{\Sigma}_\ast\right), 
 \end{align}
where $\boldsymbol{\mu}_\ast$ and $\mathbf{\Sigma}_\ast$ are defined as follows:  
\begin{equation}\begin{matrix} \label{eq:mean and covariance}
 \boldsymbol{\mu}_\ast=
 \mathbf{m}_\ast
 +\mathbf{K}_{\ast}^{T}\left(\mathbf{K}_y+\sigma_n\mathbf{I}\right)^{-1}
 \left(\mathbf{Y}-\mathbf{m}_y\right) \\
 \mathbf{\Sigma}_\ast=\mathbf{K}_{\ast\ast}+\mathbf{K}_\ast^T\left(\mathbf{K}_y+\sigma_n\mathbf{I}\right)^{-1}\mathbf{K}_\ast, \\
 \end{matrix}
\end{equation}
where $\mathbf{K}_y=k\left(\mathbf{X},\ \mathbf{X}\right),\ \mathbf{K}_{\ast\ast}=k\left(\mathbf{X}_\ast,\mathbf{X}_\ast\right)$. 
$\sigma_n\mathbf{I}$ is the error matrix having zero for all off-diagonal terms with the size $\left(M(D+1)\times N(D+1)\right)$. 
For simplicity, we denote $\mathbf{X}^{(m)}\equiv\mathbf{x}^{\left(m\right)}$ and $\left(\boldsymbol{K}\right)_{ij}=\left(k\left(\mathbf{X},\mathbf{X}^\prime\right)\right)_{ij}=k\left(\mathbf{X}^{\left(i\right)},{\mathbf{X}^\prime}^{\left(j\right)}\right)=k\left(\mathbf{x}^{\left(i\right)},{\mathbf{x}^\prime}^{\left(j\right)}\right)$. The kernel matrix can be simply represented as $\boldsymbol{K}=k\left(\mathbf{X}_\ast,\mathbf{X}\right)$. 
$\mathbf{m}_y$ and $\mathbf{m}_\ast$ are mean vectors at each site $\mathbf{X}$ and $\mathbf{X}_\ast$, respectively. 
We set the elements of the mean vector corresponding to potential data to the average of the gathered data. 
We set the elements of the mean vector corresponding to force to zero. 
The predictive Hessian $\boldsymbol{H}_\ast$ of Gaussian PES is
\begin{equation}
\boldsymbol{H}_\ast=\mathbf{K}_H^T\left(\mathbf{K}_y+\sigma_n\mathbf{I}\right)^{-1}\left(\mathbf{Y}-\mathbf{m}_y\right),
\end{equation}
where $\mathbf{K}_H=\left(\begin{matrix}\partial_{nn}\boldsymbol{K}&\partial_{mnn}\boldsymbol{K}\\\end{matrix}\right)$ is a Hessian kernel, having the second and third derivatives of covariance function as its components. 
$\partial_{nn}\boldsymbol{K}, \partial_{mnn}\boldsymbol{K}$ are tensors with shape $\left(M\times D\times D\times N\right)$ and $\left(D\times M\times D\times D\times N\right)$. 
The tensors are reshaped as the matrices of the sizes of $\left(M\times DDN\right)$ and $\left(DM\times DDN\right)$, respectively. 
Each component of tensors can be computed by
\begin{align}
    \left(\partial_{nn}\boldsymbol{K}\right)_{m,d^\prime,d,n} =&\frac{\partial}{\partial x_{d^\prime}^{\left(n\right)}}\left(\frac{\partial}{\partial x_d^{\left(n\right)}}k\left(\mathbf{x}^{\left(m\right)},\mathbf{x}^{\left(n\right)}\right)\right)\\
    \left(\partial_{mnn}\boldsymbol{K}\right)_{d,m,d^{\prime\prime},d^\prime,n} =&\frac{\partial}{\partial x_{d^{\prime\prime}}^{\left(n\right)}}\left(\frac{\partial}{\partial x_{d^\prime}^{\left(n\right)}}\left(\frac{\partial}{\partial x_d^{\left(m\right)}}k\left(\mathbf{x}^{\left(m\right)},\mathbf{x}^{\left(n\right)}\right)\right)\right),
\end{align}

Most kernels used in GP are associated with hyperparameters. 
Hyperparameters are selected to maximize the log-likelihood function:

\begin{align}
\ln{p\left(\mathbf{Y}\middle|\mathbf{X}\right)} =& \ln{\mathcal{N}\left(\mathbf{Y}\middle|\boldsymbol{m}_Y,\mathbf{K}_y\right)} \notag \\
=& -\frac{1}{2}\left(\mathbf{Y}-\boldsymbol{m}_y\right)^T\left(\mathbf{K}_y+\sigma_n\mathbf{I}\right)^{-1}\left(\mathbf{Y}-\boldsymbol{m}_y\right) \notag \\
&-\frac{1}{2}\ln{\left|\mathbf{K}_y+\sigma_n\mathbf{I}\right|}-\frac{N}{2}\ln{\left(2\pi\right).}
\label{eq:loglikelihood}
\end{align}

However, adding gradient information often breaks the positive definiteness of a kernel matrix. 
If it happens, the determinant is not guaranteed to be positive, which may make $\ln{\left|\mathbf{K}_y+\sigma_n\mathbf{I}\right|}$ impossible. 
In cases when we get a negative determinant, we generate pseudo-kernel and output vector,  $k\left(\mathbf{X}+\mathbf{dX},\ \mathbf{X}+\mathbf{dX}\right),  \mathbf{Y}+\mathbf{dY}$ where all the force information is replaced with the interpolated potential value near $\mathbf{dX}$. 
With this positive-definite guaranteed kernel, we can find hyperparameters that minimize the log-likelihood function, Eq.~(\ref{eq:loglikelihood}).

\section{Squared Exponential Kernel}
The elements of the squared exponential kernel matrices used for the MB potential are as follows: 
\begin{align}
\left(\partial_n\boldsymbol{K}\right)_{mn} =& \frac{\partial}{\partial\mathbf{x}^{\left(n\right)}}k\left(\mathbf{x}^{\left(m\right)},\mathbf{x}^{\left(n\right)}\right)
=\frac{\sigma_f}{l^2}\left(\mathbf{x}^{\left(m\right)}-\mathbf{x}^{\left(n\right)}\right)
\times k\left(\mathbf{x}^{\left(m\right)},\mathbf{x}^{\left(n\right)}\right) \\
\left(\partial_m\boldsymbol{K}\right)_{mn} =&\frac{\partial}{\partial\mathbf{x}^{\left(m\right)}}k\left(\mathbf{x}^{\left(m\right)},\mathbf{x}^{\left(n\right)}\right) =-\frac{\sigma_f}{l^2}\ \left(\mathbf{x}^{\left(m\right)}-\mathbf{x}^{\left(n\right)}\right)
\times k\left(\mathbf{x}^{\left(m\right)},\mathbf{x}^{\left(n\right)}\right)\\
\left(\partial_{mn}\boldsymbol{K}\right)_{d,m,d^\prime,n} =&\frac{\sigma_f}{l^2}\left(\delta_{dd^\prime}\ -\frac{1}{l^2}\left(\mathbf{x}_d^{\left(m\right)}-\mathbf{x}_d^{\left(n\right)}\right)\left(\mathbf{x}_{d^\prime}^{\left(m\right)}\ -\mathbf{x}_{d^\prime}^{\left(n\right)}\right)\right)
\times k\left(\mathbf{x}^{\left(m\right)},\mathbf{x}^{\left(n\right)}\right),
\end{align}
where the shape of $\partial_n\boldsymbol{K}$ is $\left(M \times 2N\right)$. 
The matrix is constructed from a $\left(M\times 2\times N\right)$ tensor and the number 2 comes from the system dimension, i.e. $D = 2$. That is, in the tensor representation, for each $m \in M$, there exists a $\left(2 \times N\right)$ matrix. 
We can reshape the $\left(2 \times N\right)$ matrix to a $2N$ dimensional array by flattening it. 
Resulting $\partial_n\boldsymbol{K}$ term forms the $\left(M\times2N\right)$ matrices. 
The $\partial_m\boldsymbol{K}$ term is constructed similarly. 
A $\left(2\times M\times N\right)$ tensor is converted to a $\left(2M\times N\right)$ matrix meaning that there exist a $\left(2 \times M\right)$ matrix for each $n\in N$.
Similarly, $\partial_{mn}\boldsymbol{K}$ is a $\left(2M \times 2N\right)$ matrix constructed from a $\left(2\times M\times2\times N\right)$ tensor. 
$\partial_{mn}\boldsymbol{K}$ consists of two parts, the global and diagonal parts. The global part is the derivative of the kernel and the diagonal part is the derivative of a coefficient. 
$\delta_{dd^\prime}$ is a Kronecker delta function having nonzero only when the terms of the second derivative are conducted in the same dimension as the first one. 

We use a zero mean Gaussian function in the MB potential model, meaning that the expectation value is $\boldsymbol{\mu}_\ast=\mathbf{K}_\ast^T\left(\mathbf{K}_y+\sigma_n\mathbf{I}\right)^{-1}\mathbf{Y}$. 
The Hessian of the Gaussian PES is calculated separately by using Hessian kernel, $\mathbf{K}_H=\left(\begin{matrix}\partial_{nn}\boldsymbol{K}&\partial_{mnn}\boldsymbol{K}\\\end{matrix}\right)$. 
The tensor representations of $\partial_{nn}\boldsymbol{K}$  and $\partial_{mnn}\boldsymbol{K}$  are given by

\begin{align}
\left(\partial_{nn}\boldsymbol{K}\right)_{m,d^\prime,d^{\prime\prime},n} =& \frac{\sigma_f}{l^2}\left\{\delta_{d^\prime d^{\prime\prime}} + \frac{1}{l^2}\left(\mathbf{x}_{d^{\prime\prime}}^{\left(m\right)} - \mathbf{x}_{d^{\prime\prime}}^{\left(n\right)}\right)\left(\mathbf{x}_{d^\prime}^{\left(m\right)}\  - \mathbf{x}_{d^\prime}^{\left(n\right)}\right) \right\} \times k\left(\mathbf{x}^{\left(m\right)}, \mathbf{x}^{\left(n\right)}\right) \\
\left(\partial_{mnn}\boldsymbol{K}\right)_{d,m,d^\prime,d^{\prime\prime},n} =& \frac{\sigma_f}{l^2}\left\{\frac{\delta_{d^\prime d^{\prime\prime}}}{l^2}\left(\mathbf{x}_d^{\left(m\right)} - \mathbf{x}_d^{\left(n\right)}\right) \right. - \frac{1}{l^4}\left(\mathbf{x}_d^{\left(m\right)} - \mathbf{x}_d^{\left(n\right)}\right)\left(\mathbf{x}_{d^{\prime\prime}}^{\left(m\right)} - \mathbf{x}_{d^{\prime\prime}}^{\left(n\right)}\right) \left(\mathbf{x}_{d^\prime}^{\left(m\right)}\ - \mathbf{x}_{d^\prime}^{\left(n\right)}\right) \notag \\
 & \left. + \frac{1}{l^2} \left(\delta_{d^{\prime\prime}d}+\delta_{d^\prime d}\right) \left(\mathbf{x}_d^{\left(m\right)} - \mathbf{x}_d^{\left(n\right)}\right)\right\} \times k\left(\mathbf{x}^{\left(m\right)}, \mathbf{x}^{\left(n\right)}\right).
\end{align}

\section{Periodic Kernel}
Suppose that $\sin\left(\mathbf{x}\right)$ is a vector that maps every element $x_i$ to $\sin{\left(x_i\right)}$. 
The kernel for AD model is
$$k\left(\mathbf{x}^{ \left( m \right)}, \mathbf{x}^{\left( n \right)} \right) = \sigma_f \exp{\left( -\frac{2}{l^2} \sum_{d}^{D}\sin^2 \frac{1}{2} \left( \mathbf{x}_d^{\left(m \right)} - \mathbf{x}_d^{\left(n\right)} \right) \right)}.$$
Tensors for each kernel matrix $\partial_n\boldsymbol{K}, \partial_m\boldsymbol{K}, \partial_{mn}\boldsymbol{K},\ \begin{matrix}\partial_{nn}\boldsymbol{K}, {\mathrm{and}} \, \partial_{mnn}\boldsymbol{K}\\
\end{matrix}$ are

\begin{equation}
\left(\partial_n\boldsymbol{K}\right)_{m,d^\prime,n}=\frac{\sigma_f}{l^2}\sin{\left(\mathbf{x}_{d^\prime}^{\left(m\right)}-\mathbf{x}_{d^\prime}^{\left(n\right)}\right)}k\left(\mathbf{x}^{\left(m\right)},\mathbf{x}^{\left(n\right)}\right)
\end{equation}

\begin{equation}
\left(\partial_m\boldsymbol{K}\right)_{d,m,n}=-\frac{\sigma_f}{l^2}\sin{\left(\mathbf{x}_d^{\left(m\right)}-\mathbf{x}_d^{\left(n\right)}\right)}k\left(\mathbf{x}^{\left(m\right)},\mathbf{x}^{\left(n\right)}\right)
\end{equation}

\begin{align}
\left(\partial_{mn}\boldsymbol{K}\right)_{d,m,d^\prime,\ n}=&\frac{\sigma_f}{l^2}\left(\frac{1}{2}\cos{2\left(\mathbf{x}_d^{\left(m\right)}-\mathbf{x}_d^{\left(n\right)}\right)}\delta_{dd^\prime} \right. \notag \\ 
- & \left. \frac{1}{l^2}\sin{\left(\mathbf{x}_{d^\prime}^{\left(m\right)}-\mathbf{x}_{d^\prime}^{\left(n\right)}\right)}\sin{\left(\mathbf{x}_d^{\left(m\right)}-\mathbf{x}_d^{\left(n\right)}\right)}\right)k\left(\mathbf{x}^{\left(m\right)},\mathbf{x}^{\left(n\right)}\right)\end{align}

\begin{align}
\left(\partial_{nn}\boldsymbol{K}\right)_{m,d^\prime,d^{\prime\prime},\ n}=&\frac{\sigma_f}{l^2}\left(\frac{1}{2}\cos{2\left(\mathbf{x}_{d^{\prime\prime}}^{\left(m\right)}-\mathbf{x}_{d^{\prime\prime}}^{\left(n\right)}\right)}\delta_{d^{\prime\prime}d^\prime}\right. \notag \\ 
+& \left. \frac{1}{l^2}\sin{\left(\mathbf{x}_{d^{\prime\prime}}^{\left(m\right)}-\mathbf{x}_{d^{\prime\prime}}^{\left(n\right)}\right)}\sin{\left(\mathbf{x}_{d^\prime}^{\left(m\right)}-\mathbf{x}_{d^\prime}^{\left(n\right)}\right)}\right)k\left(\mathbf{x}^{\left(m\right)},\mathbf{x}^{\left(n\right)}\right)
\end{align}

\begin{align}
\left(\partial_{mnn}\boldsymbol{K}\right)_{d,\ m,d^{\prime\prime},d^\prime,\ n} & = \notag \\
& \frac{\sigma_f}{l^2}\left\{-\sin{2\left(\mathbf{x}_{d^{\prime\prime}}^{\left(m\right)}-\mathbf{x}_{d^{\prime\prime}}^{\left(n\right)}\right)}\delta_{d^{\prime\prime}d^\prime}\delta_{d^{\prime\prime}d} +\frac{\delta_{d^{\prime\prime}d}}{l^2}\cos{\left(\mathbf{x}_{d^{\prime\prime}}^{\left(m\right)}-\mathbf{x}_{d^{\prime\prime}}^{\left(n\right)}\right)}\sin{\left(\mathbf{x}_{d^\prime}^{\left(m\right)}-\mathbf{x}_{d^\prime}^{\left(n\right)}\right)}\right. \notag \\
&+\frac{\delta_{d^\prime d}}{l^2}\cos{\left(\mathbf{x}_{d^\prime}^{\left(m\right)}-\mathbf{x}_{d^\prime}^{\left(n\right)}\right)}\sin{\left(\mathbf{x}_{d^{\prime\prime}}^{\left(m\right)}-\mathbf{x}_{d^{\prime\prime}}^{\left(n\right)}\right)} \notag \\
& +\frac{1}{2l^2}\cos{2\left(\mathbf{x}_{d^{\prime\prime}}^{\left(m\right)}-\mathbf{x}_{d^{\prime\prime}}^{\left(n\right)}\right)}\sin{\left(\mathbf{x}_d^{\left(m\right)}-\mathbf{x}_d^{\left(n\right)}\right)}\delta_{d^{\prime\prime}d}\delta_{d^{\prime\prime}d^\prime} \notag \\
& \left. +\frac{1}{l^4}\sin{\left(\mathbf{x}_{d^{\prime\prime}}^{\left(m\right)}-\mathbf{x}_{d^{\prime\prime}}^{\left(n\right)}\right)}\sin{\left(\mathbf{x}_{d^\prime}^{\left(m\right)}-\mathbf{x}_{d^\prime}^{\left(n\right)}\right)}\sin{\left(\mathbf{x}_d^{\left(m\right)}-\mathbf{x}_d^{\left(n\right)}\right)}\right\}k\left(\mathbf{x}^{\left(m\right)},\mathbf{x}^{\left(n\right)}\right).
\end{align}

\section{Alanine dipeptide GPAO example}
\begin{figure*}[htb!]
\begin{adjustbox}{width=1.05\textwidth}
\begin{tikzpicture}

\node[anchor=south west] (X) at (0,0){
\includegraphics[width=1\textwidth]{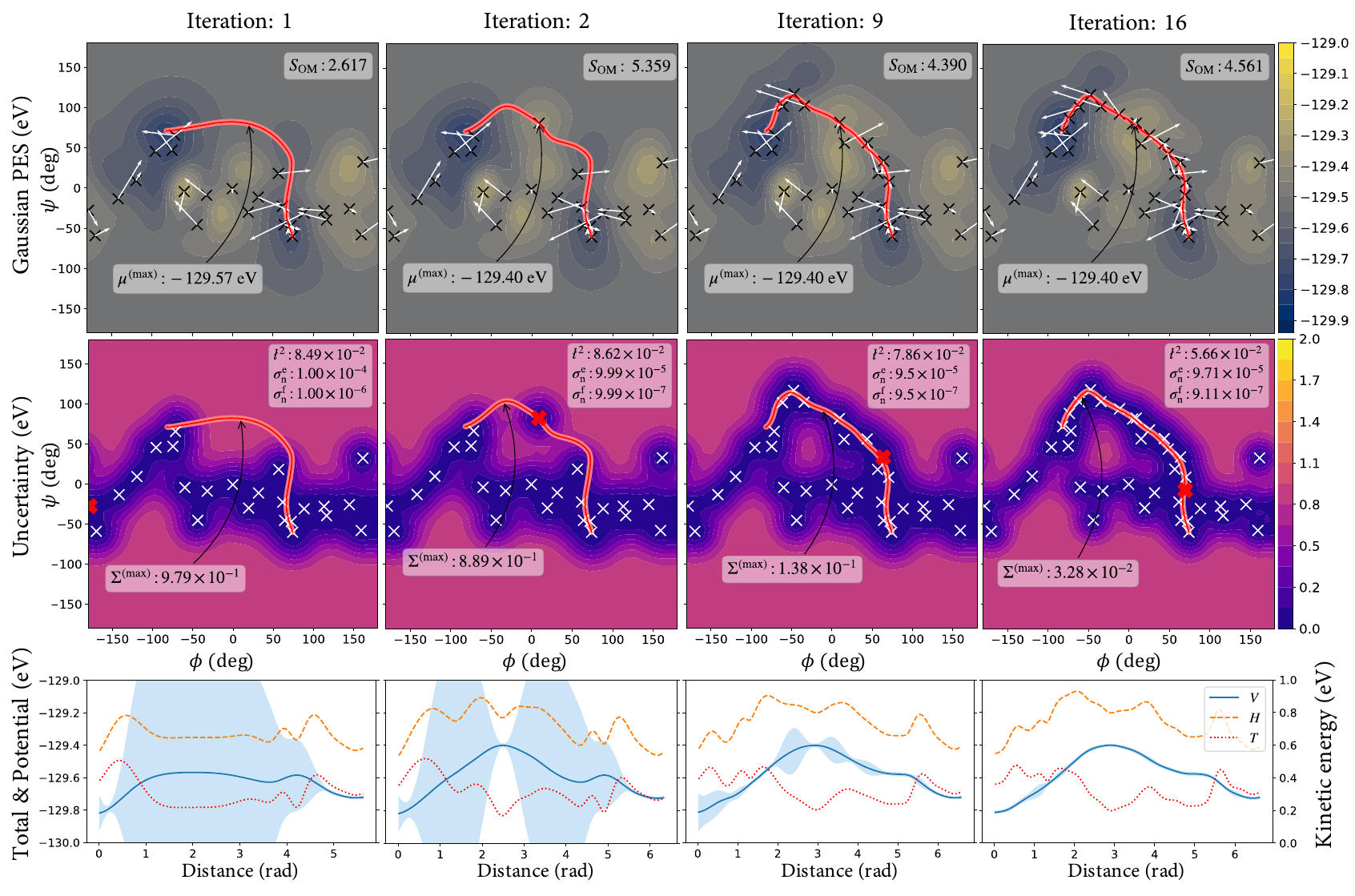}
};
\begin{scope}[x={(X.south east)},y={(X.north west)}]%

\node[text width=0in] (A) at (0.02,0.92) {(a)};
\node[text width=0in] (B) at (0.02,0.6) {(b)};
\node[text width=0in] (C) at (0.036, 0.29) {(c)};

\end{scope}%
\end{tikzpicture}
\end{adjustbox}
\caption{The progress of GP-CSA process applied to alanine dipeptide. 
(a) Gaussian PES, (b) uncertainty map and (c) energy profile along the pathway are displayed. 
Gaussian PES and uncertainty map are generated by the data marked with black (or white) crosses. 
White arrows indicate the directions and relative strengths of the data. 
The converged pathway (Iteration 16) corresponds to the pathway obtained with GP-CSA under 2 ps time step (Figure 4b-{\sffamily F}).
}
    \label{subfig:ADGPAO}
\end{figure*}

\section{Efficient global action optimization using Action-CSA}
The Action-CSA is set to start with randomly generated 40 initial pathways, followed by local optimizations using GPAO.
Afterward, we proceed with the following steps: 
\begin{enumerate}
    \item  Pick two pathways from the candidates randomly. 
    \item  Generate a new pathway from the two pathways using a crossover operator.
    \item  Apply random mutations/perturbations to the new pathway with a certain probability
    \item  Locally minimize the new pathway. 
    \item  Pick the most \textit{similar} pathway to the new pathway from the candidates. If the distance between the new pathway and a \textit{similar} pathway is lower than a cutoff distance, compare the OM action between them. Otherwise, compare the highest OM pathway among the candidates. 
    \item  If the new one has lower action than its counterpart, replace it with the new one.
    \item  Reduce the cutoff distance. If the cutoff distance is lower than a certain threshold, stop the iteration. Otherwise, go back to step 1.
\end{enumerate}

With two randomly selected pathways, we choose a random image along the pathway. 
With the selected position, we slice both trajectories at the selected image and merge them. 
After a new pathway is generated, a random mutation/perturbation is performed with a probability of 30\%.
For the pathway's mutation, the pathway is modified by adding another random number at the sine components of a pathway. 
After the local minimization of the pathway using the GPAO method, we search the nearest neighbor by measuring the Fr\'{e}chet distance between the new pathway and the current candidates. 
The nearest neighbor is considered \textit{similar}, readily accessible to each other, if the Fr\'{e}chet distance is shorter than the current cutoff distance $d_\mathrm{cut}$. 
The initial cutoff distance is set to half of the average distance between the initial pathways. 
After an iteration, we reduce $d_\mathrm{cut}$ by 2\% and repeat the whole procedure from step 1 until $d_\mathrm{cut}$ becomes below 0.05.

\section{List of parameters of molecular and surface examples}
\texttt{epoch} denotes the total transition time in a unit of $\AA \sqrt{\mathrm{u/eV}}$. \texttt{N} denotes the number of steps, and the number of sine components \texttt{Nk}, energy restraint constant $\mu_E$ as \texttt{muE}

\begin{verbatim}

Formaldehyde
Initial state (xyz) 
  1 C  5.000  5.000  5.000  F  F  F 
  2 O  3.785  5.000  4.999  T  T  T
  3 H  5.591  5.948  5.001  T  T  T
  4 H  5.591  4.052  5.001  T  T  T
Final state (xyz) 
  1 C  5.000  5.000  5.000  F  F  F 
  2 O  3.678  5.027  5.000  T  T  T
  3 H  3.387  5.963  5.000  T  T  T
  4 H  5.204  3.892  5.000  T  T  T
  
Trajectory and action parameters
 epoch                          : 5         
 N                              : 150       
 Nk                             : 100       
 gamma                          : 1.0       
 muE                            : 1.0       
 
Kernel hyperparameters bound
1.00e-03 < $\sigma_f$    < 1.00e+03
1.00e-02 < $l^2$         < 1.00e+01
1.00e-04 < $\sigma_n^e$  < 1.00e-02
1.00e-05 < $\sigma_n^f$  < 1.00e-03

VASP parameters
 prec                           : normal    
 algo                           : fast      
 nelmin                         : 4         
 ncore                          : 4         
 maxmix                         : 20        
 ismear                         : 0         
 isym                           : -1        
 istart                         : 2         
 smass                          : -2        
 encut                          : 520       
 ediff                          : 1e-05     
 xc                             : pbe       
 ivdw                           : 11        
 lcharg                         : 0         
 
Formic acid 
Initial state (xyz) 
  1 O  3.750  5.525  5.000  T  T  T 
  2 O  6.030  5.641  4.999  T  T  T 
  3 C  5.000  5.000  5.000  F  F  F 
  4 H  4.932  3.896  5.001  T  T  T 
  5 H  3.846  6.502  5.000  T  T  T 
  
Final state (xyz)
  1 O  3.784  5.615  5.000 
  2 O  6.034  5.621  5.000 
  3 C  5.000  5.000  5.000 
  4 H  4.937  3.889  5.000 
  5 H  3.078  4.939  5.000 

Action parameters
  epoch                          : 10.0    
  N                              : 150       
  Nk                             : 100       
  gamma                          : 1.0       
  muE                            : 1.0       
  
Kernel hyperparameters bound
 1.00e-03 <  $\sigma_f$   < 1.00e+03
 1.00e-02 <  $l^2$        < 1.00e+01
 1.00e-04 <  $\sigma_n^e$ < 1.00e-02
 1.00e-05 <  $\sigma_n^f$ < 1.00e-03
  
VASP Parameters
  prec                           : normal    
  algo                           : fast      
  nelmin                         : 4         
  ncore                          : 4         
  maxmix                         : 20        
  ismear                         : 0         
  isym                           : -1        
  istart                         : 2         
  smass                          : -2        
  encut                          : 520       
  ediff                          : 1e-05     
  xc                             : pbe       
  ivdw                           : 11        
  lcharg                         : 0         

Propyne
Initial state (xyz) 
  1 C  3.547  5.000  5.000  T  T  T 
  2 C  5.000  5.000  5.000  F  F  F
  3 C  6.211  5.000  5.000  T  T  T
  4 H  3.151  5.066  6.024  T  T  T
  5 H  3.152  5.854  4.430  T  T  T
  6 H  3.151  4.080  4.545  T  T  T
  7 H  7.280  5.000  5.000  T  T  T
Final state (xyz) 
  1 C  3.673  5.000  5.000  T  T  T 
  2 C  5.000  5.000  5.000  F  F  F
  3 C  6.327  5.000  5.000  T  T  T
  4 H  6.901  4.072  4.921  T  T  T
  5 H  3.100  5.929  5.080  T  T  T
  6 H  3.099  4.072  4.920  T  T  T
  7 H  6.900  5.929  5.080  T  T  T

Trajectory and action parameters
 epoch                          : 10.0      
 N                              : 150       
 Nk                             : 100       
 D                              : 18        
 gamma                          : 1.0       
 muE                            : 1.0       
 dimension                      : 18 
 
Kernel hyperparameters bound
 
 1.00e-03 < $\sigma_f$   < 1.00e+03
 1.00e-02 < $l^2$        < 1.00e+01
 1.00e-04 < $\sigma_n^e$ < 1.00e-02
 1.00e-05 < $\sigma_n^f$ < 1.00e-03
 
VASP parameters
 prec                           : normal    
 algo                           : fast      
 nelmin                         : 4         
 ncore                          : 4         
 maxmix                         : 20        
 ismear                         : 0         
 isym                           : -1        
 istart                         : 2         
 smass                          : -2        
 encut                          : 520       
 ediff                          : 1e-05     
 xc                             : pbe       
 ivdw                           : 11        
 lcharg                         : 0  
 
Gold hopping
Initial state (xyz)
  1  Al  0.000   0.000   4.000  F  F  F
  2  Al  2.864   0.000   4.000  F  F  F
  3  Al  0.000   2.864   4.000  F  F  F
  4  Al  2.864   2.864   4.000  F  F  F
  5  Al  1.432   1.432   6.025  F  F  F
  6  Al  4.296   1.432   6.025  F  F  F
  7  Al  1.432   4.296   6.025  F  F  F
  8  Al  4.296   4.296   6.025  F  F  F
  9  Al  -0.011  -0.011  8.041  T  T  T
  10 Al  2.875  -0.011   8.041  T  T  T
  11 Al  -0.011  2.875   8.041  T  T  T
  12 Al  2.875   2.875   8.041  T  T  T
  13 Au  1.432   1.432   9.775  T  T  T
Final state (xyz)
  1  Al  0.000   0.000   4.000  F  F  F
  2  Al  2.864   0.000   4.000  F  F  F
  3  Al  0.000   2.864   4.000  F  F  F
  4  Al  2.864   2.864   4.000  F  F  F
  5  Al  1.432   1.432   6.025  F  F  F
  6  Al  4.296   1.432   6.025  F  F  F
  7  Al  1.432   4.296   6.025  F  F  F
  8  Al  4.296   4.296   6.025  F  F  F
  9  Al  0.011  -0.011   8.041  T  T  T
  10 Al  2.853  -0.011   8.041  T  T  T
  11 Al  0.011   2.875   8.041  T  T  T
  12 Al  2.853   2.875   8.041  T  T  T
  13 Au  4.296   1.432   9.775  T  T  T

Trajectory and action parameters
 epoch                          : 100.0     
 N                              : 150       
 Nk                             : 50        
 D                              : 15        
 gamma                          : 1.0       
 muE                            : 1.0       
 dimension                      : 15        
 
Kernel hyperparameters bound
1.00e-03 < $\sigma_f$   < 1.00e+03
1.00e-02 < $l^2$        < 1.00e+01
1.00e-06 < $\sigma_n^e$ < 1.00e-03
1.00e-07 < $\sigma_n^f$ < 1.00e-02
 

Platinum rearrange
Initial state (xyz)
  1  Pt  -0.000  1.600  0.000  F  F  F   
  2  Pt  2.772   1.600  0.000  F  F  F   
  3  Pt  5.544   1.600  0.000  F  F  F   
  4  Pt  8.316   1.600  0.000  F  F  F   
  5  Pt  1.386   4.001  0.000  F  F  F   
  6  Pt  4.158   4.001  0.000  F  F  F   
  7  Pt  6.930   4.001  0.000  F  F  F   
  8  Pt  9.702   4.001  0.000  F  F  F   
  9  Pt  2.772   6.401  0.000  F  F  F   
  10 Pt  5.544   6.401  0.000  F  F  F   
  11 Pt  8.316   6.401  0.000  F  F  F   
  12 Pt  11.087  6.401  0.000  F  F  F   
  13 Pt  4.158   8.802  0.000  F  F  F   
  14 Pt  6.930   8.802  0.000  F  F  F   
  15 Pt  9.702   8.802  0.000  F  F  F   
  16 Pt  12.473  8.802  0.000  F  F  F   
  17 Pt  0.000   0.000  2.263  F  F  F   
  18 Pt  2.772   0.000  2.263  F  F  F   
  19 Pt  5.544   0.000  2.263  F  F  F   
  20 Pt  8.316   0.000  2.263  F  F  F   
  21 Pt  1.386   2.400  2.263  F  F  F   
  22 Pt  4.158   2.400  2.263  F  F  F   
  23 Pt  6.930   2.400  2.263  F  F  F   
  24 Pt  9.702   2.400  2.263  F  F  F   
  25 Pt  2.772   4.801  2.263  F  F  F   
  26 Pt  5.544   4.801  2.263  F  F  F   
  27 Pt  8.316   4.801  2.263  F  F  F   
  28 Pt  11.087  4.801  2.263  F  F  F   
  29 Pt  4.158   7.201  2.263  F  F  F   
  30 Pt  6.930   7.201  2.263  F  F  F   
  31 Pt  9.702   7.201  2.263  F  F  F   
  32 Pt  12.473  7.201  2.263  F  F  F   
  33 Pt  1.386   0.800  4.526  F  F  F   
  34 Pt  4.158   0.800  4.526  F  F  F   
  35 Pt  6.930   0.800  4.526  F  F  F   
  36 Pt  9.702   0.800  4.526  F  F  F   
  37 Pt  2.772   3.201  4.526  F  F  F   
  38 Pt  5.544   3.201  4.526  F  F  F   
  39 Pt  8.316   3.201  4.526  F  F  F   
  40 Pt  11.087  3.201  4.526  F  F  F   
  41 Pt  4.158   5.601  4.526  F  F  F   
  42 Pt  6.930   5.601  4.526  F  F  F   
  43 Pt  9.702   5.601  4.526  F  F  F   
  44 Pt  12.473  5.601  4.526  F  F  F   
  45 Pt  5.544   8.002  4.526  F  F  F   
  46 Pt  8.316   8.002  4.526  F  F  F   
  47 Pt  11.087  8.002  4.526  F  F  F   
  48 Pt  13.867  8.003  4.540  F  F  F   
  49 Pt  0.009   1.594  6.787  F  F  F   
  50 Pt  2.763   1.593  6.787  F  F  F   
  51 Pt  5.540   1.597  6.746  F  F  F   
  52 Pt  8.319   1.597  6.746  F  F  F   
  53 Pt  1.386   4.001  6.743  F  F  F   
  54 Pt  4.123   4.020  6.783  F  F  F   
  55 Pt  6.930   4.005  6.746  F  F  F   
  56 Pt  9.736   4.021  6.783  F  F  F   
  57 Pt  2.771   6.399  6.747  F  F  F   
  58 Pt  5.545   6.413  6.787  F  F  F   
  59 Pt  8.314   6.412  6.787  F  F  F   
  60 Pt  11.088  6.401  6.743  F  F  F   
  61 Pt  4.171   8.798  6.795  F  F  F   
  62 Pt  6.930   8.802  6.735  F  F  F   
  63 Pt  9.691   8.798  6.787  F  F  F   
  64 Pt  12.473  8.761  6.787  F  F  F   
  65 Pt  5.594   0.092  8.906  T  T  T   
  66 Pt  8.262   0.092  8.907  T  T  T   
  67 Pt  4.262   2.401  8.904  T  T  T   
  68 Pt  6.929   2.401  9.030  T  T  T   
  69 Pt  9.597   2.401  8.903  T  T  T   
  70 Pt  5.596   4.710  8.903  T  T  T   
  71 Pt  8.263   4.710  8.903  T  T  T   
Final state (xyz)
  1  Pt  -0.000  1.600  0.000  F  F  F
  2  Pt  2.772   1.600  0.000  F  F  F
  3  Pt  5.544   1.600  0.000  F  F  F
  4  Pt  8.316   1.600  0.000  F  F  F
  5  Pt  1.386   4.001  0.000  F  F  F
  6  Pt  4.158   4.001  0.000  F  F  F
  7  Pt  6.930   4.001  0.000  F  F  F
  8  Pt  9.702   4.001  0.000  F  F  F
  9  Pt  2.772   6.401  0.000  F  F  F
  10  Pt  5.544  6.401  0.000  F  F  F
  11 Pt  8.316   6.401  0.000  F  F  F
  12 Pt  11.087  6.401  0.000  F  F  F
  13 Pt  4.158   8.802  0.000  F  F  F
  14 Pt  6.930   8.802  0.000  F  F  F
  15 Pt  9.702   8.802  0.000  F  F  F
  16 Pt  12.473  8.802  0.000  F  F  F
  17 Pt  0.000   0.000  2.263  F  F  F
  18 Pt  2.772   0.000  2.263  F  F  F
  19 Pt  5.544   0.000  2.263  F  F  F
  20 Pt  8.316   0.000  2.263  F  F  F
  21 Pt  1.386   2.400  2.263  F  F  F
  22 Pt  4.158   2.400  2.263  F  F  F
  23 Pt  6.930   2.400  2.263  F  F  F
  24 Pt  9.702   2.400  2.263  F  F  F
  25 Pt  2.772   4.801  2.263  F  F  F
  26 Pt  5.544   4.801  2.263  F  F  F
  27 Pt  8.316   4.801  2.263  F  F  F
  28 Pt  11.087  4.801  2.263  F  F  F
  29 Pt  4.158   7.201  2.263  F  F  F
  30 Pt  6.930   7.201  2.263  F  F  F
  31 Pt  9.702   7.201  2.263  F  F  F
  32 Pt  12.473  7.201  2.263  F  F  F
  33 Pt  1.386   0.800  4.526  F  F  F
  34 Pt  4.158   0.800  4.526  F  F  F
  35 Pt  6.930   0.800  4.526  F  F  F
  36 Pt  9.702   0.800  4.526  F  F  F
  37 Pt  2.772   3.201  4.526  F  F  F
  38 Pt  5.544   3.201  4.526  F  F  F
  39 Pt  8.316   3.201  4.526  F  F  F
  40 Pt  11.087  3.201  4.526  F  F  F
  41 Pt  4.158   5.601  4.526  F  F  F
  42 Pt  6.930   5.601  4.526  F  F  F
  43 Pt  9.702   5.601  4.526  F  F  F
  44 Pt  12.473  5.601  4.526  F  F  F
  45 Pt  5.544   8.002  4.526  F  F  F
  46 Pt  8.316   8.002  4.526  F  F  F
  47 Pt  11.087  8.002  4.526  F  F  F
  48 Pt  13.867  8.003  4.540  F  F  F
  49 Pt  0.009   1.594  6.787  F  F  F
  50 Pt  2.763   1.593  6.787  F  F  F
  51 Pt  5.540   1.597  6.746  F  F  F
  52 Pt  8.319   1.597  6.746  F  F  F
  53 Pt  1.386   4.001  6.743  F  F  F
  54 Pt  4.123   4.020  6.783  F  F  F
  55 Pt  6.930   4.005  6.746  F  F  F
  56 Pt  9.736   4.021  6.783  F  F  F
  57 Pt  2.771   6.399  6.747  F  F  F
  58 Pt  5.545   6.413  6.787  F  F  F
  59 Pt  8.314   6.412  6.787  F  F  F
  60 Pt  11.088  6.401  6.743  F  F  F
  61 Pt  4.171   8.798  6.795  F  F  F
  62 Pt  6.930   8.802  6.735  F  F  F
  63 Pt  9.691   8.798  6.787  F  F  F
  64 Pt  12.460  8.765  6.776  F  F  F
  65 Pt  5.575   0.106  8.922  T  T  T
  66 Pt  8.180   0.076  8.872  T  T  T
  67 Pt  4.219   2.424  8.933  T  T  T
  68 Pt  6.924   2.417  8.977  T  T  T
  69 Pt  8.199   4.749  8.862  T  T  T
  70 Pt  2.902   4.716  8.871  T  T  T
  71 Pt  5.548   4.747  8.934  T  T  T
 
Trajectory and action parameters
 epoch                          : 100.0     
 N                              : 150       
 Nk                             : 50        
 gamma                          : 1.0       
 muE                            : 1.0       
 
Kernel hyperparameters bound
1.00e-03 < $\sigma_f$   < 1.00e+03
1.00e-02 < $l^2$        < 1.00e+01
1.00e-06 < $\sigma_n^e$ < 1.00e-02
1.00e-07 < $\sigma_n^f$ < 1.00e-03
 
Platinum diffusion
Initial state (xyz)
  1   Pt  -0.000  1.600   0.000  F  F  F 
  2   Pt  2.772   1.600   0.000  F  F  F
  3   Pt  5.544   1.600   0.000  F  F  F
  4   Pt  8.316   1.600   0.000  F  F  F
  5   Pt  11.087  1.600   0.000  F  F  F
  6   Pt  13.859  1.600   0.000  F  F  F
  7   Pt  1.386   4.001   0.000  F  F  F
  8   Pt  4.158   4.001   0.000  F  F  F
  9   Pt  6.930   4.001   0.000  F  F  F
  10  Pt  9.702   4.001   0.000  F  F  F
  11  Pt  12.473  4.001   0.000  F  F  F
  12  Pt  15.245  4.001   0.000  F  F  F
  13  Pt  2.772   6.401   0.000  F  F  F
  14  Pt  5.544   6.401   0.000  F  F  F
  15  Pt  8.316   6.401   0.000  F  F  F
  16  Pt  11.087  6.401   0.000  F  F  F
  17  Pt  13.859  6.401   0.000  F  F  F
  18  Pt  16.631  6.401   0.000  F  F  F
  19  Pt  4.158   8.802   0.000  F  F  F
  20  Pt  6.930   8.802   0.000  F  F  F
  21  Pt  9.702   8.802   0.000  F  F  F
  22  Pt  12.473  8.802   0.000  F  F  F
  23  Pt  15.245  8.802   0.000  F  F  F
  24  Pt  18.017  8.802   0.000  F  F  F
  25  Pt  5.544   11.202  0.000  F  F  F
  26  Pt  8.316   11.202  0.000  F  F  F
  27  Pt  11.087  11.202  0.000  F  F  F
  28  Pt  13.859  11.202  0.000  F  F  F
  29  Pt  16.631  11.202  0.000  F  F  F
  30  Pt  19.403  11.202  0.000  F  F  F
  31  Pt  6.930   13.603  0.000  F  F  F
  32  Pt  9.702   13.603  0.000  F  F  F
  33  Pt  12.473  13.603  0.000  F  F  F
  34  Pt  15.245  13.603  0.000  F  F  F
  35  Pt  18.017  13.603  0.000  F  F  F
  36  Pt  20.789  13.603  0.000  F  F  F
  37  Pt  0.001   0.017   2.210  F  F  F
  38  Pt  2.774   0.017   2.211  F  F  F
  39  Pt  5.545   0.018   2.210  F  F  F
  40  Pt  8.317   0.019   2.209  F  F  F
  41  Pt  11.089  0.017   2.209  F  F  F
  42  Pt  13.860  0.016   2.210  F  F  F
  43  Pt  1.387   2.404   2.215  F  F  F
  44  Pt  4.159   2.406   2.216  F  F  F
  45  Pt  6.931   2.407   2.214  F  F  F
  46  Pt  9.703   2.407   2.215  F  F  F
  47  Pt  12.474  2.404   2.216  F  F  F
  48  Pt  15.245  2.403   2.216  F  F  F
  49  Pt  2.774   4.802   2.208  F  F  F
  50  Pt  5.546   4.804   2.208  F  F  F
  51  Pt  8.319   4.805   2.208  F  F  F
  52  Pt  11.090  4.804   2.209  F  F  F
  53  Pt  13.861  4.802   2.208  F  F  F
  54  Pt  16.632  4.801   2.208  F  F  F
  55  Pt  4.161   7.195   2.218  F  F  F
  56  Pt  6.934   7.197   2.218  F  F  F
  57  Pt  9.706   7.197   2.218  F  F  F
  58  Pt  12.478  7.197   2.219  F  F  F
  59  Pt  15.249  7.195   2.218  F  F  F
  60  Pt  18.020  7.194   2.218  F  F  F
  61  Pt  5.549   9.584   2.196  F  F  F
  62  Pt  8.322   9.586   2.198  F  F  F
  63  Pt  11.094  9.588   2.201  F  F  F
  64  Pt  13.862  9.587   2.200  F  F  F
  65  Pt  16.634  9.584   2.197  F  F  F
  66  Pt  19.407  9.584   2.196  F  F  F
  67  Pt  6.934   12.001  2.181  F  F  F
  68  Pt  9.706   12.003  2.182  F  F  F
  69  Pt  12.477  12.000  2.186  F  F  F
  70  Pt  15.248  12.003  2.181  F  F  F
  71  Pt  18.020  12.002  2.180  F  F  F
  72  Pt  20.792  12.001  2.180  F  F  F
  73  Pt  1.389   0.835   4.485  F  F  F
  74  Pt  4.161   0.837   4.488  F  F  F
  75  Pt  6.935   0.841   4.483  F  F  F
  76  Pt  9.707   0.838   4.484  F  F  F
  77  Pt  12.477  0.834   4.487  F  F  F
  78  Pt  15.247  0.833   4.487  F  F  F
  79  Pt  2.773   3.214   4.436  F  F  F
  80  Pt  5.546   3.216   4.435  F  F  F
  81  Pt  8.319   3.218   4.435  T  T  T
  82  Pt  11.091  3.216   4.436  T  T  T
  83  Pt  13.860  3.213   4.436  F  F  F
  84  Pt  16.631  3.212   4.436  F  F  F
  85  Pt  4.161   5.571   4.438  F  F  F
  86  Pt  6.934   5.573   4.438  T  T  T
  87  Pt  9.708   5.574   4.438  T  T  T
  88  Pt  12.479  5.576   4.438  T  T  T
  89  Pt  15.248  5.570   4.438  F  F  F
  90  Pt  18.019  5.569   4.438  F  F  F
  91  Pt  5.551   7.958   4.501  F  F  F
  92  Pt  8.324   7.961   4.502  T  T  T
  93  Pt  11.091  7.963   4.496  T  T  T
  94  Pt  13.875  7.963   4.495  T  T  T
  95  Pt  16.640  7.960   4.501  T  T  T
  96  Pt  19.409  7.958   4.500  F  F  F
  97  Pt  6.938   10.394  4.443  F  F  F
  98  Pt  9.711   10.397  4.445  F  F  F
  99  Pt  12.484  10.393  4.458  F  F  F
  100 Pt  15.258  10.399  4.441  F  F  F
  101 Pt  18.027  10.397  4.441  F  F  F
  102 Pt  20.797  10.394  4.441  F  F  F
  103 Pt  8.324   12.824  4.436  F  F  F
  104 Pt  11.097  12.826  4.440  F  F  F
  105 Pt  13.867  12.831  4.437  F  F  F
  106 Pt  16.641  12.828  4.435  F  F  F
  107 Pt  19.412  12.824  4.434  F  F  F
  108 Pt  22.182  12.822  4.435  F  F  F
  109 Pt  4.168   8.906   6.617  F  F  F
  110 Pt  6.938   8.907   6.620  T  T  T
  111 Pt  9.699   8.911   6.624  T  T  T
  112 Pt  12.489  8.852   6.655  T  T  T
  113 Pt  15.276  8.917   6.624  F  F  F
  114 Pt  18.031  8.911   6.619  F  F  F
  115 Pt  5.552   11.218  6.669  F  F  F
  116 Pt  8.314   11.224  6.671  T  T  T
  117 Pt  11.083  11.232  6.697  T  T  T
  118 Pt  13.895  11.239  6.686  T  T  T
  119 Pt  16.662  11.234  6.668  F  F  F
  120 Pt  19.420  11.223  6.668  F  F  F
  121 Pt  6.936   13.535  6.603  F  F  F
  122 Pt  9.700   13.551  6.598  F  F  F
  123 Pt  12.485  13.547  6.603  F  F  F
  124 Pt  15.273  13.561  6.595  F  F  F
  125 Pt  18.037  13.544  6.599  F  F  F
  126 Pt  20.802  13.534  6.602  F  F  F
  127 Pt  12.499  10.385  8.657  T  T  T
Final state (xyz)
  1   Pt  -0.000  1.600   0.000  F  F  F
  2   Pt  2.772   1.600   0.000  F  F  F
  3   Pt  5.544   1.600   0.000  F  F  F
  4   Pt  8.316   1.600   0.000  F  F  F
  5   Pt  11.087  1.600   0.000  F  F  F
  6   Pt  13.859  1.600   0.000  F  F  F
  7   Pt  1.386   4.001   0.000  F  F  F
  8   Pt  4.158   4.001   0.000  F  F  F
  9   Pt  6.930   4.001   0.000  F  F  F
  10  Pt  9.702   4.001   0.000  F  F  F
  11  Pt  12.473  4.001   0.000  F  F  F
  12  Pt  15.245  4.001   0.000  F  F  F
  13  Pt  2.772   6.401   0.000  F  F  F
  14  Pt  5.544   6.401   0.000  F  F  F
  15  Pt  8.316   6.401   0.000  F  F  F
  16  Pt  11.087  6.401   0.000  F  F  F
  17  Pt  13.859  6.401   0.000  F  F  F
  18  Pt  16.631  6.401   0.000  F  F  F
  19  Pt  4.158   8.802   0.000  F  F  F
  20  Pt  6.930   8.802   0.000  F  F  F
  21  Pt  9.702   8.802   0.000  F  F  F
  22  Pt  12.473  8.802   0.000  F  F  F
  23  Pt  15.245  8.802   0.000  F  F  F
  24  Pt  18.017  8.802   0.000  F  F  F
  25  Pt  5.544   11.202  0.000  F  F  F
  26  Pt  8.316   11.202  0.000  F  F  F
  27  Pt  11.087  11.202  0.000  F  F  F
  28  Pt  13.859  11.202  0.000  F  F  F
  29  Pt  16.631  11.202  0.000  F  F  F
  30  Pt  19.403  11.202  0.000  F  F  F
  31  Pt  6.930   13.603  0.000  F  F  F
  32  Pt  9.702   13.603  0.000  F  F  F
  33  Pt  12.473  13.603  0.000  F  F  F
  34  Pt  15.245  13.603  0.000  F  F  F
  35  Pt  18.017  13.603  0.000  F  F  F
  36  Pt  20.789  13.603  0.000  F  F  F
  37  Pt  0.001   0.017   2.210  F  F  F
  38  Pt  2.774   0.017   2.211  F  F  F
  39  Pt  5.545   0.018   2.210  F  F  F
  40  Pt  8.317   0.019   2.209  F  F  F
  41  Pt  11.089  0.017   2.209  F  F  F
  42  Pt  13.860  0.016   2.210  F  F  F
  43  Pt  1.387   2.404   2.215  F  F  F
  44  Pt  4.159   2.406   2.216  F  F  F
  45  Pt  6.931   2.407   2.214  F  F  F
  46  Pt  9.703   2.407   2.215  F  F  F
  47  Pt  12.474  2.404   2.216  F  F  F
  48  Pt  15.245  2.403   2.216  F  F  F
  49  Pt  2.774   4.802   2.208  F  F  F
  50  Pt  5.546   4.804   2.208  F  F  F
  51  Pt  8.319   4.805   2.208  F  F  F
  52  Pt  11.090  4.804   2.209  F  F  F
  53  Pt  13.861  4.802   2.208  F  F  F
  54  Pt  16.632  4.801   2.208  F  F  F
  55  Pt  4.161   7.195   2.218  F  F  F
  56  Pt  6.934   7.197   2.218  F  F  F
  57  Pt  9.706   7.197   2.218  F  F  F
  58  Pt  12.478  7.197   2.219  F  F  F
  59  Pt  15.249  7.195   2.218  F  F  F
  60  Pt  18.020  7.194   2.218  F  F  F
  61  Pt  5.549   9.584   2.196  F  F  F
  62  Pt  8.322   9.586   2.198  F  F  F
  63  Pt  11.094  9.588   2.201  F  F  F
  64  Pt  13.862  9.587   2.200  F  F  F
  65  Pt  16.634  9.584   2.197  F  F  F
  66  Pt  19.407  9.584   2.196  F  F  F
  67  Pt  6.934   12.001  2.181  F  F  F
  68  Pt  9.706   12.003  2.182  F  F  F
  69  Pt  12.477  12.000  2.186  F  F  F
  70  Pt  15.248  12.003  2.181  F  F  F
  71  Pt  18.020  12.002  2.180  F  F  F
  72  Pt  20.792  12.001  2.180  F  F  F
  73  Pt  1.389   0.835   4.485  F  F  F
  74  Pt  4.161   0.837   4.488  F  F  F
  75  Pt  6.935   0.841   4.483  F  F  F
  76  Pt  9.707   0.838   4.484  F  F  F
  77  Pt  12.477  0.834   4.487  F  F  F
  78  Pt  15.247  0.833   4.487  F  F  F
  79  Pt  2.773   3.214   4.436  F  F  F
  80  Pt  5.546   3.216   4.435  F  F  F
  81  Pt  8.307   3.202   4.439  T  T  T
  82  Pt  11.093  3.197   4.472  F  F  F
  83  Pt  13.860  3.213   4.436  F  F  F
  84  Pt  16.631  3.212   4.436  F  F  F
  85  Pt  4.161   5.571   4.438  F  F  F
  86  Pt  6.914   5.570   4.433  T  T  T
  87  Pt  9.681   5.585   4.461  T  T  T
  88  Pt  12.515  5.591   4.456  T  T  T
  89  Pt  15.248  5.570   4.438  F  F  F
  90  Pt  18.019  5.569   4.438  F  F  F
  91  Pt  5.551   7.958   4.501  F  F  F
  92  Pt  8.308   7.973   4.497  T  T  T
  93  Pt  11.090  7.967   4.505  T  T  T
  94  Pt  13.875  7.984   4.495  T  T  T
  95  Pt  16.639  7.964   4.500  T  T  T
  96  Pt  19.409  7.958   4.500  F  F  F
  97  Pt  6.938   10.394  4.443  F  F  F
  98  Pt  9.711   10.397  4.445  F  F  F
  99  Pt  12.484  10.393  4.458  F  F  F
  100 Pt  15.258  10.399  4.441  F  F  F
  101 Pt  18.027  10.397  4.441  F  F  F
  102 Pt  20.797  10.394  4.441  F  F  F
  103 Pt  8.324   12.824  4.436  F  F  F
  104 Pt  11.097  12.826  4.440  F  F  F
  105 Pt  13.867  12.831  4.437  F  F  F
  106 Pt  16.641  12.828  4.435  F  F  F
  107 Pt  19.412  12.824  4.434  F  F  F
  108 Pt  22.182  12.822  4.435  F  F  F
  109 Pt  4.168   8.906   6.617  F  F  F
  110 Pt  6.928   8.901   6.624  T  T  T
  111 Pt  9.702   8.895   6.622  T  T  T
  112 Pt  12.476  8.898   6.620  T  T  T
  113 Pt  15.276  8.917   6.624  F  F  F
  114 Pt  18.031  8.911   6.619  F  F  F
  115 Pt  5.552   11.218  6.669  F  F  F
  116 Pt  8.314   11.210  6.677  T  T  T
  117 Pt  11.086  11.210  6.676  T  T  T
  118 Pt  13.859  11.214  6.676  T  T  T
  119 Pt  16.662  11.234  6.668  F  F  F
  120 Pt  19.420  11.223  6.668  F  F  F
  121 Pt  6.936   13.535  6.603  F  F  F
  122 Pt  9.700   13.551  6.598  F  F  F
  123 Pt  12.485  13.547  6.603  F  F  F
  124 Pt  15.273  13.561  6.595  F  F  F
  125 Pt  18.037  13.544  6.599  F  F  F
  126 Pt  20.802  13.534  6.602  F  F  F
  127 Pt  11.110  4.843   6.433  T  T  T
  
Trajectory and action parameters
 epoch                          : 100.0     
 N                              : 150       
 Nk                             : 50        
 D                              : 48        
 gamma                          : 1.0       
 muE                            : 1.0       
 dimension                      : 48        
 
Kernel hyperparameters bound
1.00e-03 < $\sigma_f$   < 1.00e+03
1.00e-02 < $l^2$        < 1.00e+01
1.00e-06 < $\sigma_n^e$ < 1.00e-02
1.00e-07 < $\sigma_n^f$ < 1.00e-03

\end{verbatim}